\documentclass[aoas,MSNbibl,nameyear,nameyear,rotating,dvips]{arximspdf}
\usepackage{mathbh}
\usepackage{dcolumn}
\usepackage{multirow}
\usepackage{graphicx}

\doi{10.1214/12-AOAS623} %kopijuoti is PTS
\volume{7}
\issue{2}
\pubyear{2013}
\firstpage{1111}
\lastpage{1138}

\makeatletter

\newcolumntype{d}[1]{D{.}{.}{#1}}

\newcommand{\mathbbm}{\mathbh}

\newcommand{\iiint}{\int\!\!\int\!\!\int}

\newcommand{\bbr}{\mathbb{R}}
\newcommand{\Ga}{{\Gamma}}

\makeatother

\begin{document}
\begin{frontmatter}

\title{A Bayesian linear model for the high-dimensional inverse
problem of seismic tomography}
\runtitle{A Bayesian linear model for seismic tomography}

\begin{aug}
\author[A]{\fnms{Ran} \snm{Zhang}\corref{}\thanksref{t1}\ead[label=e1]{ran.zhang@ma.tum.de}},
\author[A]{\fnms{Claudia} \snm{Czado}\ead[label=e2]{cczado@ma.tum.de}}
\and
\author[B]{\fnms{Karin} \snm{Sigloch}\ead[label=e3]{karin.sigloch@geophysik.uni-muenchen.de}}
\runauthor{R. Zhang, C. Czado and K. Sigloch}
\affiliation{Technische Universit\"at M\"unchen, Technische Universit\"at
M\"unchen and Ludwig-Maximilians Universit\"at M\"unchen}
\address[A]{R. Zhang\\
C. Czado\\
Center for Mathematical Sciences\\
Technische Universit\"at M\"unchen\\
Boltzmannstrasse 3\\
85748 Garching bei M\"unchen\\
Germany\\
\printead{e1}\\
\hphantom{E-mail: }\printead*{e2}} %adresu isvedimo komanda gale!
\address[B]{K. Sigloch\\
Department of Earth\\
\quad and Environmental Sciences\\
Ludwig-Maximilians-Universit\"{a}t\\
Theresienstrasse 41\\
80333 Munich\\
Germany\\
\printead{e3}}
\end{aug}

\thankstext{t1}{Supported by the Munich Center of Advanced Computing
(MAC/IGSSE) and by the LRZ Supercomputing Center.}

% HISTORY:
\received{\smonth{3} \syear{2012}}
\revised{\smonth{9} \syear{2012}}

% ABSTRACT
%
\begin{abstract}
We apply a linear Bayesian model to seismic tomography, a
high-dimensional inverse problem in geophysics. The objective is to
estimate the three-dimensional structure of the earth's interior from
data measured at its surface. Since this typically involves estimating
thousands of unknowns or more, it has always been treated as a
linear(ized) optimization problem. Here we present a Bayesian
hierarchical model to estimate the joint distribution of earth
structural and earthquake source parameters. An ellipsoidal spatial
prior allows to accommodate the layered nature of the earth's mantle.
With our efficient algorithm we can sample the posterior distributions
for large-scale linear inverse problems and provide precise uncertainty
quantification in terms of parameter distributions and credible
intervals given the data. We apply the method to a full-fledged
tomography problem, an inversion for upper-mantle structure under
western North America that involves more than 11,000 parameters. In
studies on simulated and real data, we show that our approach retrieves
the major structures of the earth's interior as well as classical
least-squares minimization, while additionally providing uncertainty
assessments.
\end{abstract}

% KEYWORDS

\end{frontmatter}

%s1 #&#
\section{Introduction}\label{sec1}
% overview
Seismic tomography is a geophysical imaging method that allows to
estimate the three-dimensional structure of the earth's deep
interior, using observations of seismic waves made at its surface.
Seismic waves generated by moderate or large earthquakes travel
through the entire planet, from crust to core, and can be recorded
by seismometers anywhere on earth. They are by far the most highly
resolving wave type available for exploring the interior at depths
to which direct measurement methods will never penetrate (tens to
thousands of kilometers). Seismic tomography takes the shape of a
large, linear(ized) inverse problem, typically featuring thousands
to millions of measurements and similar numbers of parameters to
solve for.

To first order, the earth's interior is layered under the
overwhelming influence of gravity. Its resulting, spherically
symmetric structure had been robustly estimated by the 1980s
[\citet{dziewonskianderson1981,kennettengdahl1991}] and is
characterized by $O(10^2)$ parameters. Since then seismologists
have been mainly concerned with estimating lateral deviations from
this spherically symmetric reference model [\citet{nolet2008}].
Though composed of solid rock, the earth's mantle is in constant
motion (the mantle extends from roughly 30 km to 2900 km depth and
is underlain by the fluid iron core). Rock masses are rising and
sinking at velocities of a few centimeters per year, the
manifestation of advective heat transfer: the hot interior slowly
loses its heat into space. This creates slight lateral variations in
material properties, on the order of a few percent, relative to the
statically layered reference model. The goal of seismic tomography
is to map these three-dimensional variations, which embody the
dynamic nature of the planet's interior.

Beneath well-instrumented regions---such as our chosen example, the
United States---seismic waves are capable of resolving mantle
heterogeneity on scales of a few tens to a few hundreds of
kilometers. Parameterizing the three-dimensional earth, or even just
a small part of it, into blocks of that size results in the
mentioned large number of unknowns, which mandate a linearization of
the inverse problem. Fortunately this is workable, thanks to the
rather weak lateral material deviations of only a few percent
(larger differences cannot arise in the very mobile mantle).

Seismic tomography is almost always treated as an optimization
problem. Most often a least squares approach is followed requiring
general matrix inverses
[\citet
{akietal1976,crosson1976,montellietal2004,siglochmcquarrienolet2008}],
while adjoint techniques are used when an explicit matrix
formulation is computationally too expensive
[\citet{tromptapeliu2005,sieminskiliu2007,fichtneretal2009}].
While probabilistic seismic tomography using Markov chain Monte Carlo
(MCMC) methods has been given considerable attention by the geophysical
(seismological) community, these applications have been restricted to
linear or nonlinear problems of much lower dimensionality assuming
Gaussian errors [\citet
{mosegaardtarantola1995,mosegaardtarantola2002,sambridgemosegaard2002}].
%%\citep{bodin:sambridge:2009,debski:2010,khanetal:2011,mosca:2012}.
%random walk proposal and apply it to the nonlinear inversion of
%gravity data. \citet{sambridgemosegaard2002} summarize Monte Carlo
%techniques such as the neighborhood algorithm, simulated annealing and
%other sampling approaches, with various geophysical applications.
For example, \citet{debski2010} compares the damped least-squares
method (LSQR), a genetic algorithm and the Metropolis--Hastings (MH)
algorithm in a low-dimensional linear tomography problem involving
copper mining data. He finds that the MCMC sampling technique
provides more robust estimates of velocity parameters compared to
the other approaches. \citet{bodinsambridge2009} capture the
uncertainty of the velocity parameters in a linear model by
selecting the representation grid of the corresponding field, using
a reversible jump MCMC (RJMCMC) approach. In
Bodin et~al. (\citeyear{bodinetalJGR2012,bodinetalGJI2012}) again RJMCMC algorithms
are developed to solve certain transdimensional nonlinear tomography
problems with Gaussian errors, assuming unknown variances.
\citet{khanetal2011} and \citet{mosca2012} study seismic and
thermo-chemical structures of the lower mantle and solve a
corresponding low-dimensional nonlinear problem using a standard
MCMC algorithm.

For exploring high-dimensional parameter space the MCMC sampling
faces difficulties in evaluating the expensive nonlinear physical
model while efficiently traversing the high-dimensional parameter
space. We approach linearized tomographic problems (physical forward
model inexpensive to solve) in a Bayesian framework, for a fully
dimensioned, continental-scale study that features $\approx$53,000
data points and $\approx$11,000 parameters. To our knowledge, this
is by far the highest dimensional application of Monte Carlo
sampling to a seismic tomographic problem so far. Assuming Gaussian
distributions for the error and the prior, our MCMC sampling scheme
allows for characterization of the posterior distribution of the
parameters by incorporating flexible spatial priors using Gaussian
Markov random field (GMRF). Spatial priors using GMRF
arise in spatial statistics
[\citet{pettitt2002,congdon2003,rue2005}], where they are mainly
used to model spatial correlation. In our geophysical context we
apply a spatial prior to the parameters rather than to the error
structure, since the parameters represent velocity anomalies in
three-dimensional space. Thanks to the sparsity of the linearized
physical forward matrix as well as the spatial prior sampling from
the posterior density, a high-dimensional multivariate Gaussian can be
achieved by a Cholesky decomposition technique from
\citet{wilkinson2002} or \citet{rue2005}. Their technique is
improved by using a different permutation algorithm. To demonstrate
the method, we estimate a three-dimensional model of mantle
structure, that is, variations in seismic wave velocities, beneath the
Unites States down to 800 km depth.

%% Applied work in other disciplines
Our approach is also applicable to other kinds of travel time
tomography, such as cross-borehole tomography or mining-induced
seismic tomography [\citet{debski2010}]. Other types of tomography,
such as X-ray tomography in medical imaging, can also be recast as a
linear matrix problem of large size with a very sparse forward
matrix. However, the response is measured on pixel areas and, thus,
the error structure is governed by a spatial Markov random field,
while the regression parameters are modeled nonspatially using, for
example, Laplace priors [\citet{kolehmainenetal2007,djafari2012}].
Some other inverse problems such as image deconvolution and computed
tomography [\citet{bardsley2012}], electromagnetic
source problems deriving from electric and magnetic encephalography,
cardiography [\citet{hamalainen1994,uulelaetal1999,kaipio2007}] or
convection-diffusion contamination transport problems
[\citet{flathetal2011}] can also be written as linear models.
However, the physical forward matrix of those problems is dense in
contrast to the situation we consider. For solutions to these
problems, matrix-inversion or low-rank approximation to the
posterior covariance matrix, as introduced in \citet{flathetal2011},
are applied to high-dimensional linear problems. In image
reconstruction problems \citet{bardsley2012} demonstrates Gibbs
sampling on (1D and 2D-) images using an intrinsic GMRF prior with the
preconditioned conjugate gradient method in cases where efficient
diagonalization or Cholesky decomposition of the posterior covariance
matrix is not available. In other tomography problems, such as
electrical capacitance tomography, electrical
impedance tomography or optical absorbtion and scattering
tomography, the physical forward model cannot be linearized, so that
the Bayesian treatment of those problems is limited to
low dimensions [\citet{kaipio2007,watzenigfox2009}].
%
%Other linear(ized) physical inverse problems such as
%convection-diffusion problems
%covariance matrix (as the $\Omega_{\beta}$ defined in
%linear inverse problems in convection-diffusion problems. The
%fundamental difference between their physical model and ours is that
%the forward matrix (the $X$ matrix in our case) is dense whereas ours
%is sparse. That is why they developed an approximation technique to
%simulate the posterior covariance matrix while we show in our paper
%that due to sparsity of the physical forward matrix direct decomposing
%the posterior precision matrix is efficient for the sampling of high
%dimensional model space.

The remainder of this paper is organized as follows: Section \ref{sec2}
describes the geophysical forward model and the seismic travel time
data. Section \ref{sec3} discusses flexible specifications for the spatial
prior of the three-dimensional velocity model and the
Metropolis--Gibbs sampling algorithm for estimating its posterior
distribution. Method performance under various model assumptions is
examined in simulation studies in Section \ref{sec4}. Section \ref{sec5} applies the
method to real travel time data, which have previously been used in
conventional tomography [\citet
{siglochmcquarrienolet2008,sigloch2010}], allowing for comparison.
Section \ref{sec6} discusses the
advantages, limitations and possible extensions of our model.

\section{Geophysical models and the data}\label{sec2}
Here we explain the physics and the data that enter seismic
tomography and how they are formulated into a linear inverse
problem, which will be treated by our Markov chain Monte Carlo
method in subsequent sections.

%s2.1 #&#
\subsection{The linear inverse problem of seismic tomography}\label{sec2.1}
Every larger earthquake generates seismic waves of sufficient
strength to be recorded by seismic stations around the globe. Such
seismograms are time series at discrete surface locations, that is,
spatially sparse point samples of a continuous wavefield that exists
%
%f1 #&#
\begin{figure}

\includegraphics{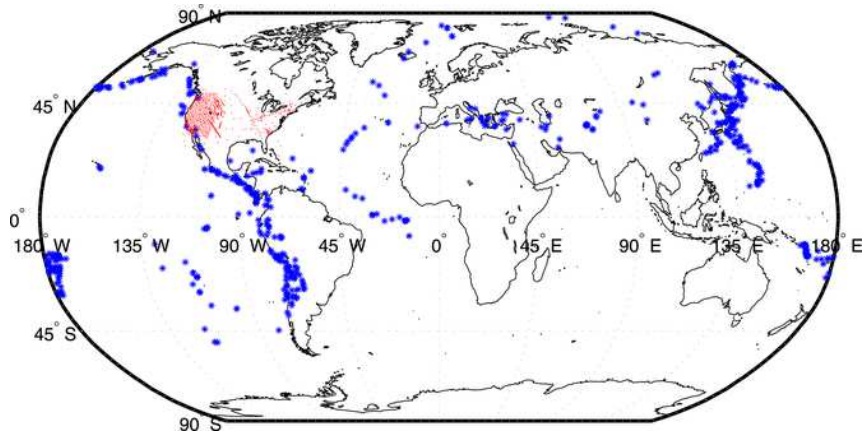}

\caption{Distribution of the seismic wave sources (large earthquakes,
blue) and receivers (seismic broadband stations, red) that generated
our data. This is a regional tomography study that includes only data
recorded in North America. In the mantle under this region, down to a
few hundreds of kilometers depth, paths of incoming waves cross densely
and from many directions, yielding good resolution for a
three-dimensional imaging study.}
\label{allevtsstats}
\end{figure}
everywhere inside the earth and at its surface. Figure~\ref{allevtsstats}
illustrates the spatial distribution of sources
(large earthquakes, blue) and receivers (seismic broadband stations,
red) that generated our data. Each datum $y_{i}$ measures the
difference between an observed arrival time
$y_{i}^{\mathrm{obs}}$ of a seismic wave $i$ and its predicted
arrival time $y_{i}^{\mathrm{pred}}$:
\[
y_{i} = y_{i}^{\mathrm{obs}} - y_{i}^{\mathrm{pred}}.
\]
$y_{i}^{\mathrm{pred}}$ is evaluated using the spherically
symmetric reference model IASP91 by \citet{kennettengdahl1991}. For
the teleseismic P waves used in our application, this difference
$y_i$ would typically be on the order of one second, whereas
$y_{i}^{\mathrm{obs}}$ and $y_{i}^{\mathrm{pred}}$ are on
the order 600--1000 seconds. $y_i$ can be explained by slightly
decreasing the modeled velocity in certain sub-volumes of the
mantle.

%f2 #&#
\begin{figure}

\includegraphics{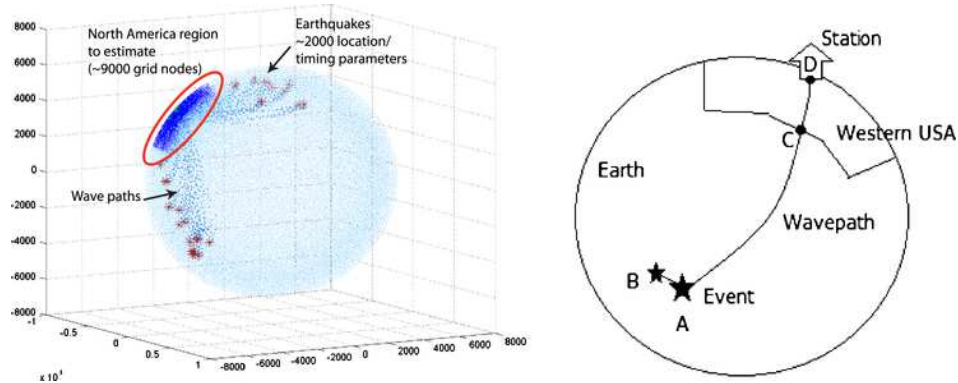}

\caption{Physical setup and forward modeling of the seismic tomography
problem. Left: parametrization of the spherical earth. Grid nodes are
shown as blue dots. The goal is to estimate seismic velocity deviations
$\bolds{\beta}$ at $\sim$9000 grid nodes under North America, inside
the subvolume marked by the red ellipse. Red stars mark a few of the
earthquake sources shown in Figure \protect\ref{allevtsstats}. The
densified point clouds, between the sources and a few stations in North
America, map out the sensitivity kernels of the selected wave paths.
Each sensitivity kernel fills one row of matrix $X$. Left: schematic
illustration of the components of an individual wave path.}
\label{fwmat}
\end{figure}

We adopt the parametrization and a subset of the data measured by
\citet{siglochmcquarrienolet2008}. The earth is meshed as a sphere
of irregular tetrahedra with 92,175 mesh nodes.
%Since this dimension is currently not manageable for MCMC sampling, we
%regard as free parameters only 8977 mesh nodes that are located
%beneath the western U.S., i.e. between latitudes $20^{\circ}$N to
%$60^{\circ}$N, longitudes $90^{\circ}$W to $130^{\circ}$W, and 0--800km
%depth and tetrahedra nodes are spaced by 60--150km.
%The parameters of interests on the mesh nodes are the relative
%velocity deviation of the western U.S. upper mantle with respect to
%the reference Earth velocity model IASP91
%given set of mesh nodes $M_{\mathrm{usa}}$ located beneath the western
%U.S. as a subset of mesh node $M_{Earth}$ of the entire earth.
At each mesh mode, the parameters of interest are the relative
velocity variation of the mantle with respect to the reference
velocity of spherically-symmetric model IASP91
[\citet{kennettengdahl1991}]. The parameter vector is denoted as
$\bolds{\beta}:=(\beta(\mathbf{r}),\mathbf{r}\in
M_{\mathrm{Earth}})\in\bbr^{92{,}175}$, where the set of mesh node
$M_{\mathrm{Earth}}$ fills the entire interior of the earth.
Since both travel time deviations $y_{i}$ and the $\beta(\mathbf{r})$
are small, the wave equation may be linearized around the layered
reference model:
%
%e1 #&#
\begin{equation}
\label{phy1a} y_{i} = \iiint_{\mathrm{Earth}} x_{i}(\mathbf{r})
\beta(\mathbf{r}) \,d^3\mathbf{r},
\end{equation}
where $x_{i}(\mathbf{r})\in\bbr$ represents the Fr\'{e}chet sensitivity
kernel of the $i$th wavepath, that is, the partial derivatives of the
chosen misfit measure or data $y_{i}$ with respect to the parameters
$\beta(\mathbf{r})$. After numerical integration of kernel $x_{i}(\mathbf
{r})$ onto the
mesh, (\ref{phy1a})~takes the form
%
%e2 #&#
\begin{equation}
\label{phy1b} y_{i} = \sum_{\mathbf{r}\in M_{\mathrm{Earth}}}x_{i}(
\mathbf{r})\beta(\mathbf{r}) = \mathbf{x}_{i}'\bolds{\beta}.
\end{equation}
Geometrically speaking, row vector $\mathbf{x}_{i}'$ maps out the
mantle subvolume that would influence the travel time $y_{i}$ if some
velocity anomaly $\beta(\mathbf{r})$ were located within it. This
sensitivity region between an earthquake and a station essentially
has ray-like character (Figure \ref{fwmat}), though in physically
more sophisticated approximations, the ray widens into a banana
shape [\citet{dahlenhungnolet2000}]. Over the past decade, intense
research effort has gone into the computability of sensitivity
kernels under more and more realistic approximations
[\citet{dahlenhungnolet2000,tromptapeliu2005,tian2007b,nolet2008}].
Since this issue is only tangential to our focus, we
chose to keep the sensitivity calculations as simple as possible by
modeling them as rays (the $\mathbf{x}_{i}'$ are computed only once and
stored). We note that the dependence of $x_{i}$ on $\bolds{\beta}$ can
be neglected, as is common practice. This is justified
by two facts: (i) velocity anomalies $\bolds{\beta}$ deviate from those
of the
(spherically symmetric) reference model by only a few percent, since
the very mobile mantle does not support larger disequilibria, and
(ii), even though the ray path in the true earth differs (slightly)
from that in the reference model, this variation affects the
travel time observable only to second order, according to Fermat's
principle [and analogous arguments for true finite-frequency
sensitivities,
\citet{dahlenhungnolet2000,nolet2008,merceratnolet2012}].
Whatever the exact modeling is, it is very sparse, since
every ray or banana visits only a small subvolume of the entire
mantle---this sparsity is important for the computational
efficiency of the MCMC sampling.

Gathering all $N$ observations, (\ref{phy1b}) can be rewritten as
$\mathbf{y}=X\bolds{\beta}$, where sparse matrix $X\in\bbr^{N\times d}$
contains in its rows the $N$ sensitivity kernels. The left panel of
Figure \ref{fwmat} illustrates the sensitivity kernels between one
station and several earthquakes (i.e., several matrix rows). In
practice, the problem never attains full rank, so that
regularization must be added to remove the remaining nonuniqueness.
The linear system $\mathbf{y}=X\bolds{\beta}$ is usually solved by some
sparse matrix solver---a popular choice is the Sparse Equations and
Least Squares (LSQR) algorithm by \citet{paigesaunders1982}, which
minimizes $\|X\bolds{\beta}-\mathbf{y}\|^2 + \lambda^2 \|\bolds{\beta}\|^2$, where
$\lambda$ is a regularization parameter that removes the
underdeterminacy in $X$
[\citet
{montellietal2004,siglochmcquarrienolet2008,tiansiglochnolet2009,debski2010}].

In summary, we have formulated the seismic tomography problem as it
is overwhelmingly practiced by the geophysical community today. We
use travel time differences $y_{i}$ as the misfit criterion, that is, as
input data to the inverse problem, and seek to estimate the
three-dimensional distribution of seismic velocity deviations
$\bolds{\beta}$ that have caused these travel time anomalies. The
sensitivity kernels $\mathbf{x}_{i}'$ are modeled using ray theory, a
high-frequency approximation to the full wave equation. In the
conventional optimization approach, a regularization term is added,
and the inverse problem is solved by minimizing the L2 norm misfit.

%s2.2 #&#
\subsection{Setup of our example problem}\label{sec2.2}
Since all 92,175 velocity deviation parameters of the entire earth
are currently not manageable for MCMC sampling, we regard as free
parameters only 8977 of those parameters which are located beneath
the western U.S., that is, between latitudes $20^{\circ}$N to
$60^{\circ}$N, longitudes $90^{\circ}$W to $130^{\circ}$W, and
0--800 km depth. Tetrahedra nodes are spaced by 60--150 km. We denote
this subset of velocity parameters as~$\bolds{\beta}_{\mathrm{usa}}$.

Besides velocity parameters, we also consider the uncertainty in the
location and the origin time of each earthquake source, which
contribute to the travel time measurement. Government and research
institutions routinely publish location estimates for every larger
earthquake, but any event may easily be mistimed by a few seconds,
and mislocated by ten or more kilometers (corresponding to a travel
duration of 1 s or more). This is a problem, since the structural
heterogeneities themselves only generate travel time delays on the
order of a few seconds. Hence, the exact locations and timings of the
earthquakes---or rather: their deviations from the published
catalogue values---need to be treated as additional free
parameters, to be estimated jointly with the structural parameters.
These so-called ``source corrections'' are captured by
three-dimensional shift corrections of the hypocenter
($\bolds{\beta}_{\mathrm{hyp}}$) and time corrections
($\bolds{\beta}_{\mathrm{time}}$) per earthquake.

Using the LSQR method, \citet{siglochmcquarrienolet2008} jointly
estimate all 92,175 parameters together with these ``source
corrections.'' Using those LSQR solutions,\vadjust{\goodbreak} we have two modeling
alternatives for the earth structural inversion with $N$ travel
delay time observations $\mathbf{y}\in\bbr^{N}$:
%
%e3 #&#
\begin{equation}
\label{stat1} \mbox{\textit{Model} 1:}\quad\mathbf{y}_{\mathrm{usa}} =
X_{\mathrm{usa}}\bolds{\beta}_{\mathrm{usa}} + \bolds{\varepsilon},\qquad \bolds
{\varepsilon}\sim
\mathcal{N}_N\biggl(\mathbf{0}, \frac{1}{\phi}I_N
\biggr),
\end{equation}
where $X_{\mathrm{usa}}\in\bbr^{N\times8977}$ denotes the
ensemble of sensitivity kernels of the western USA.
$\mathcal{N}_{N}(\bolds{\mu}, \Sigma)$ denotes the $N$-dimensional
multivariate normal distribution with mean $\bolds{\mu}$ and
covariance $\Sigma$, and the $N$-dimensional unity matrix is denoted
by $I_N$. In model 1, we only estimate the velocity parameters
$\bolds{\beta}_{\mathrm{usa}}$ and keep the part of the travel
delay time for the corrections parameters (path AB in right panel of
Figure~\ref{fwmat}) fixed at the LSQR solutions of
$\bolds{\beta}_{\mathrm{hyp}}$ and
$\bolds{\beta}_{\mathrm{time}}$ estimated by
\citet{siglochmcquarrienolet2008}. The extended model with joint
estimation of source corrections is given by
%
%e4 #&#
\begin{eqnarray}
\label{stat2} \mbox{\textit{Model} 2:}\quad\mathbf{y}_{\mathrm{cr}} &=&
X_{\mathrm{usa}}\bolds{\beta}_{\mathrm{usa}} + X_{\mathrm{hyp}} \bolds{
\beta}_{\mathrm{hyp}} + X_{\mathrm{time}}\bolds{\beta}_{\mathrm
{time}} +
\bolds{
\varepsilon},\nonumber\\[-8pt]\\[-8pt]
\bolds{\varepsilon}&\sim&\mathcal{N}_N\biggl(\mathbf{0},
\frac{1}{\phi}I_N\biggr).\nonumber
\end{eqnarray}
Here we apply the travel delay time $\mathbf{y}_{\mathrm{cr}}$
assuming that the part of the travel time running through path AC is
given. This given part of the travel times is again based on the
LSQR solution estimated by \citet{siglochmcquarrienolet2008}.
%%%%%%%
%We can also write \eqref{stat2} as $\mathbf{y}_{\mbox{\tiny{cr}}} = X\bolds{
%$X:= \left(X_{\mathrm{usa}}, X_{\mathrm{hyp}}, X_{\mathrm{time}}
%for $d_{\mbox{\tiny{cr}}}:=d_{\mathrm{usa}}+d_{\mathrm{hyp}}+d_{

The number of travel time data from source-receiver pairs is
$N=53\mbox{,}270$, collected from 760 stations and 529 events. The number of
hypocenter correction parameters is 1587 (529 earthquakes${}\times{}$3)
and there are 529 time correction parameters. \citet{sigloch2008}
found that in the uppermost mantle, between 0 km to 100 km depth, the
velocity can deviate by more than $\pm5\%$ from the spherically
symmetric reference model. As depth increases, the mantle becomes more
homogeneous and the velocity deviates less from the reference model.

%%\left(
%%\begin{array}{llll}
%%X_{usa} & X_{rest} & X_{hyp} & X_{time}\\
%%\end{array}
%%\right)
%or
%+ X_{hyp}\bolds{\beta}_{hyp} + X_{time}\bolds{\beta}_{time},
% \begin{minipage}{1.0\linewidth}
% \includegraphics[scale=0.4]{./graphics/normQQplot_traveltime.jpg}
% %\end{minipage}
% \hfill
% %\begin{minipage}{1.0\linewidth}
% \includegraphics[scale=0.4]{./graphics/histogram.pdf}
% \end{minipage}
% \caption{QQ-plot and histogram of the travel time data. Dashed line:
%theoretical density of the normal
%distribution with mean 1; long dashed line: Student's
%$t$-distributions with
%location 1 and df 3, dot dashed line: skew normal
%distribution with location 2 and skewness -1; solid line: skew
%$t$-distribution with location 2, skewness -1 and df 3.}

% \subsection{Spatial modeling using the conditional autoregressive
%model}
%s3 #&#
\section{Estimation method}\label{sec3}
%s3.1 #&#
\subsection{Modeling the spatial structure of the velocity parameters}\label{sec3.1}
\begin{figure}

\includegraphics{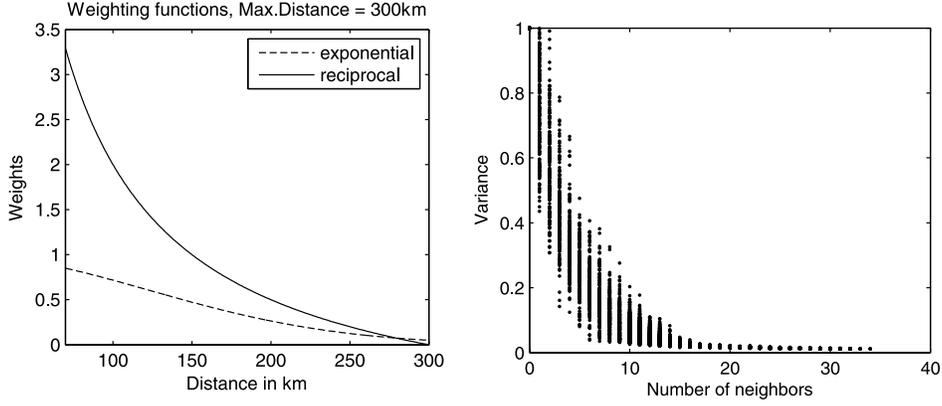}

\caption{Left: exponential and reciprocal weight functions for the
spatial prior, for $D=150$ km and $D=300$ km. Right: the trade-off
relationship between numbers of neighbors and the prior variance
$\operatorname{diag}(Q^{-1}(\psi)),\psi=10$, $D = 150$ km, $w =
\mbox{reciprocal}$ weights.}
\label{weightsfunc}
\end{figure}

In both models (\ref{stat1}) and (\ref{stat2}) we have the spatial
parameter $\bolds{\beta}_{\mathrm{usa}}$, which we denote
generically as $\bolds{\beta}$ in this section. In the Bayesian
approach we need a proper prior distribution for this high-dimensional
parameter vector $\bolds{\beta}$. To account for their
spatially correlated structure, we apply the conditional
autoregressive model (CAR) and assume a Markov random field
structure for $\bolds{\beta}$. This assumption says that the
conditional distribution of the local characteristics $\beta_i$,
given all other parameters $\beta_j$, $j\neq i$, only depends on the
neighbors, that is,
$P(\beta_i\mid\bolds{\beta}_{-i})=P(\beta_i\mid\beta_j,j\sim i)$,
where
$\bolds{\beta}_{-i}:=(\beta_1,\ldots,\beta_{i-1},\beta_{i+1},\ldots,\beta_d)'$
and ``\mbox{$\sim$}$i$'' denotes the set of neighbors of site~$i$. The CAR
model and its application have been investigated in many studies,
such as \citet{pettitt2002} or \citet{rue2005}.\vadjust{\goodbreak} Since the earth is
heterogeneous and layered, lateral correlation length scales are
larger than over depths, and so we propose an ellipsoidal
neighborhood structure for the velocity parameters. Let
$(x_j,y_j,z_j)'\in\bbr^3$ be the positions of the $i$th and the
$j$th nodes in Cartesian coordinates. The $j$th node is a neighbor
of node $i$ if the ellipsoid equation is satisfied, that is,
$ (\frac{x_i-x_j}{D_x} )^2+ (\frac
{y_i-y_j}{D_y} )^2+ (\frac{z_i-z_j}{D_z} )^2\leq1$.
To add a rotation of the ellipsoid to an arbitrary direction in the
space, we could simply modify the vector $(x_i-x_j, y_i-y_j,
z_i-z_j)'$ to $R(x_i-x_j, y_i-y_j, z_i-z_j)'$ with a rotation matrix
$R = R_x R_y R_z$ for given rotation matrices in the $x$, $y$ and $z$
directions, respectively. The spherical neighborhood structure is a
special case of the ellipsoidal structure with $D_x = D_y = D_z$.
Let $D$ be the maximum distance of $D_x$, $D_y$ and~$D_z$. For
weighting the neighbors we adopt either the exponential $w_e(\cdot)$
or reciprocal weight functions $w_r(\cdot)$, that is,
%
%e5 #&#
\begin{equation}
\label{wfuncs} w_e(d_{ij}):= \exp\biggl\{-
\frac{3d_{ij}^2}{D^2}\biggr\} \quad\mbox{and}\quad w_r(d_{ij}):=
\frac{D}{d_{ij}} - 1,
\end{equation}
%
%w_e(d_{ij}) = \exp\left\{-\frac{3d_{ij}^2}{D^2}\right\},
%and the reciprocal weight function proposed by \citet{pettitt2002}
%w_r(d_{ij}):= \frac{D}{d_{ij}} - 1,
where $d_{ij}$ is the Euclidean distance between node $i$ and node
$j$. The exponential weight function is bounded while the reciprocal
weight function is unbounded. Those weighting functions have been
studied by \citet{pettitt2002} or \citet{congdon2003}. The left
panel of Figure \ref{weightsfunc} illustrates the weight functions
for $D=300$ km.

%
%f3 #&#

Let $\omega(d_{ij})$ be either $w_e(\cdot)$ or $w_r(\cdot)$ in
(\ref{wfuncs}). To model the spatial structure of
$\bolds{\beta}_{\mathrm{usa}}$ in (\ref{stat1}) and
(\ref{stat2}), a CAR model is used. Following \citet{pettitt2002},
let
$\bolds{\beta}_{\mathrm{usa}}\sim\mathcal{N}_{d_{\mathrm
{usa}}} (\mathbf{0}, \frac{1}{\eta_{\mathrm{usa}}}
Q^{-1}(\psi) )$ with precision matrix
%%\sim\mathcal{N}\left(\sum_{j:j\sim i}c_{ij}\beta_{j}, \frac{m_{ii}}{
%with
%&c_{ij}=\left\{
%0 & j = i\\
%&m_{ii}:=\frac{1}{1+|\psi|\sum_{i:k\sim i}\omega(d_{ik})}, i=1,\ldots,d.
%
%e6 #&#
\begin{equation}
\label{qmat} %&c_{ij}=\left\{
%0 & j = i\\
%&m_{ii}:=\frac{1}{1+|\psi|\sum_{i:k\sim i}\omega(d_{ik})}, i=1,\ldots,d.\\
Q_{ij}(\psi):=\cases{\displaystyle 1+|\psi|\sum
_{i:j\sim i}\omega(d_{ij}), &\quad $i = j$,
\vspace*{2pt}\cr
\displaystyle -\psi
\omega(d_{ij}), &\quad $i \neq j, i \sim j \mbox{ for } \psi\in\bbr$.}
\end{equation}
%
%C)^{-1}M\right),
They showed that $Q$ is symmetric and positive definite, and that
conditional correlations can be explicitly determined.
% between $\beta_i$
%and $\beta_j$ is given all other parameter $\bolds{\beta}_{-\{i,j\}}$ is
%Cor[\beta_i,\beta_j\mid\bolds{\beta}_{-\{i,j\}}] = \frac{\psi
For $\psi\rightarrow0$, the precision matrix $Q$ converges to the
identity matrix, that is, $\psi= 0$ corresponds to independent
elements of $\bolds{\beta}$. The precision matrix in (\ref{qmat}) for
both elliptical and spherical cases indicates anisotropic covariance
structure and depends on the distance between nodes, the number of
neighbors of each node and the weighting functions. The elliptical
precision matrix additionally depends on the orientation. The right
panel of Figure \ref{weightsfunc} shows the trade-off between
numbers of neighbors and prior variance, which indicates that the
more neighbors the $i$th node has, the smaller is its prior variance
$(Q^{-1}(\psi))_{ii}$. Posterior distribution of velocity parameters
from regions with less neighborhood information can be rough, since
they are not highly regularized due to the large prior covariances.
This may produce sharp edges in the tomographic image. However, this
is a more realistic modeling method since one is more sure about the
optimization solution if a velocity parameter has more neighbors.
Moreover, this prior specification is adapted to the construction of
the tetrahedral mesh: regions with many nodes have better ray
coverage than regions with less nodes. In summary, the prior
incorporates diverse spatial knowledge about the velocity
parameters. Since a precision matrix is defined, which is sparse and
positive definite, it provides a computational advantage in sampling
from a high-dimensional Gaussian distribution as required in our
algorithm (shown in the following sections).

%However in the Bayesian framework credible interval could be used for
%finding the most confident regions for the solutions. Usually the
%region
%with less confidence as bigger credible interval, one cannot use the
%posterior mode solution as the
%optimal solution.
%For the parameter $\psi$ we chose
%$\psi\sim\mathcal{N}(\mu_{\psi},\sigma_{\psi}^2)$ to
%Define $Q:=M^{-1}(I_d-C)$ (precision) or $Q^{-1}:=(I_d-C)^{-1}M$
%(covariance) where $C=(c_{ij})_{i,j=1,\ldots,d}$ with $c_{ii}=0$,
%$c_{ij}\neq0$ if $i\sim j$, $i\neq j$, and $M:=\diag(m_{ii}),
%i=1,\ldots,d$ is a diagonal matrix such that $(I_d-C)^{-1}M$ is
%symmetric and positive definite.
%It satisfies $c_{ij}m_{jj}=c_{ji}m_{ii}$, $i,j=1,\ldots,d$.
%The corresponding pdf is then
%p(\psi)\propto(\sigma_{\psi}^2)^{-\frac{1}{2}}\exp\left\{-\frac{1}{2
%Furthermore, $\omega(\cdot)$ is defined as in \eqref{dist1} and
%$\bolds{\beta}\mid

%s3.2 #&#
\subsection{A Gibbs--Metropolis sampler for parameter estimation in
high dimensions}\label{sec3.2}
% Anfang der section immer eine zusammenfassung, worum es in den
%sections geht
To quantify uncertainty, we adopt a Bayesian approach. Posterior
inference for the model parameters is facilitated by a Metropolis
within Gibbs sampler [\citet{brooksetal2011}]].
%Pioneer work on Bayesian treatment of regression models with normal or
%more general
%distributions such as elliptical distributions can be found in
%this topic can be seen in \citet{sahu2003},
%
%Pioneer work on Bayesian treatment of regression models with normal or
%more general
%distributions such as elliptical distributions can be found in
%this topic can be seen in \citet{kimmallick2003}, \citet{sahu2003},
%
%Because the physical
%characteristics of the travel time arrivals at the seismic stations
%across western USA, we adopt the CAR model to model the spatially
%correlated structure of the travel time data. As the regularization
%prior for $\bolds{\beta}$ we also use the CAR model to describe the 3D
%neighborhood structure of the discrete grid nodes.
Recall the linear model in (\ref{stat2}),
\[
\mathbf{Y} = X\bolds{\beta} + \bolds{\varepsilon},\qquad \bolds{\varepsilon}\sim
\mathcal{N}_N\biggl(\mathbf{0}, \frac{1}{\phi}I_N
\biggr),
\]
where\vspace*{1pt}
$\bolds{\beta}:=(\bolds{\beta}_{\mathrm{usa}},\bolds{\beta}_{\mathrm{hyp}},
\bolds{\beta}_{\mathrm{time}})'$
and $X:= (X_{\mathrm{usa}}, X_{\mathrm{hyp}},
X_{\mathrm{time}} )$. We now specify the prior
distribution of $\bolds{\beta}$ as
\[
\label{betaprior} \bolds{\beta} \sim\mathcal{N}_{d} (\bolds{
\beta}_0, \Sigma_{\beta} ) \qquad\mbox{with }\bolds{
\beta}_0:=(\bolds{\beta}_{0,\mathrm{usa}},\bolds{\beta}_{0,\mathrm{hyp}},
\bolds{\beta}_{0,\mathrm{time}})'.
\]
The prior covariance matrix $\Sigma_{\beta}$ is chosen as
%
%e7 #&#
\begin{equation}
\label{sigmabeta} \Sigma_{\beta}:= \pmatrix{
\displaystyle \frac{1}{\eta_{\mathrm{usa}}}Q^{-1}(\psi) & \mathbf{0} & \mathbf{0}
\vspace*{2pt}\cr
\mathbf{0} &
\displaystyle \frac{1}{\eta_{\mathrm{hyp}}}I_{d_{\mathrm{hyp}}} & \mathbf{0}
\vspace*{2pt}\cr
\mathbf{0} & \mathbf{0} &
\displaystyle \frac{1}{\eta_{\mathrm{time}}}I_{d_{\mathrm
{time}}}}.
\end{equation}
Since we are interested in modeling positive spatial dependence, we
impose that the spatial dependence parameter $\psi$ is the truncated
normal distribution a priori, that is,
$\psi\sim\mathcal{N}(\mu_{\psi},\sigma^2_{\psi})\mathbbm
{1}(\psi>0)$.
The priors for the precision scale parameters $\eta_{\mathrm{usa}}$,
$\eta_{\mathrm{hyp}}$, $\eta_{\mathrm{time}}$ and $\phi$ are specified
in terms of a Gamma distribution $\Gamma(a,b)$ with density
$g(x; a,b)=\frac{b^a}{\Gamma(a)}x^{a-1}\exp\{-bx\}$, $x>0$. The
corresponding first two moments are $\frac{a}{b}$ and $\frac{a}{b^2}$,
respectively.

The MCMC procedure is derived as follows: the full conditionals of
$\bolds{\beta}$ are
%
%e8 #&#
\begin{eqnarray}
\label{betafullcond}
\bolds{\beta}\mid\mathbf{y},\psi,\bolds{\eta}
&\sim&
\mathcal{N}_d\bigl(\Omega_{\beta}^{-1}\bolds{
\xi}_{\beta}, \Omega_{\beta
}^{-1}\bigr),
\nonumber\\[-8pt]\\[-8pt]
&&\eqntext{\mbox{with }\Omega_{\beta}:=\Sigma_{\beta
}^{-1} + \phi
X'X, \bolds{\xi}_{\beta}:=\Sigma_{\beta}^{-1}
\bolds{\beta}_0 + \phi X'\mathbf{y}}
\end{eqnarray}
and
$\bolds{\eta}:=(\eta_{\mathrm{usa}},\eta_{\mathrm{hyp}},\eta
_{\mathrm{time}})$.
For $\eta_{\mathrm{usa}}$, $\eta_{\mathrm{hyp}}$,
$\eta_{\mathrm{time}}$ and $\phi$, the full conditionals are
again Gamma distributed.
%
%$\Omega_{\beta}:=\Sigma_{\beta}^{-1} + \phi X'X$, $\bolds{\xi}_{\beta}:=
% with $\Omega_{\beta}:=\Sigma_{\beta}^{-1} + \phi X'Q_yX$, $\bolds{\xi}_{
% with $\Omega_{\beta}:=\eta Q_{\beta}(\psi) + \phi X'Q_y(\psi_y)X$, $
%&\eta_{usa} \mid\bolds{\beta},\psi,\mathbf{y} \sim\Gamma\left(a_{\eta,usa}+
%&\eta_{hyp} \mid\bolds{\beta},\mathbf{y} \sim\Gamma\left(a_{\eta,hyp}+
%&\eta_{time}\mid\bolds{\beta},\mathbf{y} \sim\Gamma\left(a_{\eta,time}+
%&\phi\mid\mathbf{y},\bolds{\beta} \sim\Ga\left(a_{\phi} + \frac{N}{2}, b_{
%
%with $\Omega_{\beta}:=\eta Q_{\beta}(\psi) + \phi X'Q_y(\psi_y)X$, $
% %&\propto\eta^{a_{\eta}+\frac{d}{2}-1}\exp\left\{-\eta\left(b_{\eta}+
The estimation of $\psi$ requires a Metropolis--Hastings (MH) step.
The logarithm of the full conditional of $\psi$ is proportional to
\begin{eqnarray*}
%(1) &\psi\sim\mathcal{N}(\mu_{\psi}, \sigma_{\psi}^2), \psi
\log\pi(\psi\mid\mathbf{y}, \bolds{\beta}, \bolds{\eta})
&\propto&\frac{1}{2}\log\bigl|Q(\psi)\bigr|\\
&&{} - \frac{\eta_{\mathrm
{usa}}}{2}(\bolds{
\beta}_{\mathrm{usa}} - \bolds{\beta}_{0,\mathrm{usa}})'Q(\psi)
(\bolds{
\beta}_{\mathrm{usa}} - \bolds{\beta}_{0,\mathrm{usa}})\\
&&{} - \frac
{(\psi- \mu
_{\psi})^2}{2\sigma_{\psi}^2}.
\end{eqnarray*}
For the MH step, we choose a truncated normal random walk proposal
for $\psi$ to obtain a new sample, that is,
$\mathcal{N}(\psi^{\mathrm{old}}, \bar{\sigma}_{\psi}^2)\mathbbm
{1}(\psi>0)$.
%{
%& \mbox{Sample } \psi^{\star} \sim q(\cdot|\psi^{\mbox{old}})
%& \mbox{Accept } \psi^{\mbox{new}}:=\psi^{\star} \mbox{ if }
We use a Cholesky decomposition with permutation to obtain a sample
of $\bolds{\beta}$ in (\ref{betafullcond}) (Section~\ref{sec3.4}). The method
by \citet{pettitt2002}, solving a sparse matrix equation, is not
useful. Here, computing the determinant of the Cholesky factor of
$Q(\psi)$ is much more efficient than calculating its eigenvalues,
due to the size and sparseness of $Q(\psi)$.
%Finally we choose fixed hyper-parameters: $a_{\phi}$, $b_{\phi}$ for $
%$a_{\eta,\mathrm{usa}}$, $b_{\eta,\mathrm{usa}}$ for $\eta_{usa}$; $a_{
%$a_{\eta,\mathrm{time}}$, $b_{\eta,\mathrm{time}}$ for $\eta_{
%and $\mu_{\psi}$, $\sigma_{\psi}^2$ for $\psi$.

%s3.3 #&#
\subsection{Relationship to ridge regression}\label{sec3.3}
To show the relationship between our approach and ridge regression
(also called Tikhonov regularization), we consider only model 1. For
simplicity we neglect the notation ``usa'' in (\ref{stat1}). The
analysis is also applicable to model 2.

Let $\hat{\bolds{\beta}}{}^{\mathrm{ridge}}(\lambda):=(X'X+\lambda
I_d)^{-1}X'\mathbf{y}$ be the corresponding ordinary ridge regression
(ORR) estimate with shrinkage parameter $\lambda$
[\citet{hoerlkennard1970,swindel1976,debski2010}]. For given
hyperparameters $\eta$, $\phi$ and $\psi$, the full conditional of
$\bolds{\beta}$ is $\bolds{\beta}\mid\eta,\phi,\psi\sim
\mathcal{N}_{d}(\Omega^{-1}_{\beta}\bolds{\xi}_{\beta},\Omega
^{-1}_{\beta})$
with $\Omega_{\beta}:=\eta Q(\psi)+\phi X'X$ and
$\bolds{\xi}_{\beta}:=\eta Q(\psi)\bolds{\beta}_{0}+\phi X'\mathbf{y}$. The
corresponding full conditional mean can therefore be expressed as
\[
E[\bolds{\beta}\mid\mathbf{y},\psi,\eta]= \biggl(X'X + \frac{\eta}{\phi}
Q(\psi) \biggr)^{-1} \biggl(X'y + \frac{\eta}{\phi} Q(
\psi) \bolds{\beta}_{0} \biggr).
\]
This is close to the modified ridge regression estimator
$\hat{\bolds{\beta}}{}^{\mathrm{ridge}}(\lambda,\bolds{\beta}_{0}):=
(X'X + \lambda I_d)^{-1}(X'\mathbf{y} + \lambda\bolds{\beta}_0)$ defined
in \citet{swindel1976}. We can see that if $\psi\rightarrow0$, then
$\frac{\eta}{\phi} Q(\psi) \rightarrow\frac{\eta}{\phi}$, which is
the equivalent to $\lambda$ in the modified ridge regression. This
shows that the prior precision matrix $\eta Q(\psi)$ is a
regularization matrix with parameter $\psi$ controlling the prior
covariance. As discussed in Section \ref{sec3.1}, the prior covariance
$\frac{1}{\eta}Q^{-1}(\psi)$ also varies with the specified weights
in (\ref{wfuncs}) with maximum distance $D$ and with number of
neighboring nodes. For large $\psi$ or large weights function
values, as well as large number of neighbors, the prior variances
are small, which well reflects the prior knowledge about the data
coverage and parameter uncertainty. Thus, the full conditional mean
is close to the prior mean in this case.

\subsection{Computational issues}\label{sec3.4}
Since the size of the travel time data requires high-dimensional
parameters to be estimated, the traditional method of sampling the
parameter vector $\bolds{\beta}$ from
$\mathcal{N}_{d}(\Omega_{\beta}^{-1}\bolds{\xi}_{\beta}, \Omega
_{\beta}^{-1})$
directly, as defined in (\ref{betafullcond}), is not efficient with
respect to computing time. We instead use a Cholesky decomposition
of $\Omega_{\beta}$. Since the sensitivity kernel $X$ is sparse, and
the prior covariance matrix is sparse and positive definite, the matrix
$\Omega_{\beta}$ remains sparse and symmetric positive definite.
Therefore, we can reduce the cost of the Cholesky decompositions.
%The Cholesky decomposition usually requires $n^3/3$ FLOPS for a dense
%matrix.
%Since our matrix is symmetric and sparse we can reduce the cost of the
%Cholesky decomposition of a large matrix.
For this we apply an approximate minimum degree ordering algorithm
(AMD algorithm) to find a permutation $P$ of $\Omega_{\beta}$ so
that the number of nonzeros in its Cholesky factor is reduced
[\citet{amestoydavisduff1996}]. In our case, the number of
nonzeros of the full conditional precision matrix $\Omega_{\beta}$
in (\ref{betafullcond}) is about $5\%$ of all elements. After this
permutation the nonzeros of the Cholesky factor are reduced by
$50\%$ compared to the original number of nonzeros.

To sample a multivariate normal distributed vector after permutation,
we follow \citet{rue2005}. Given the permutation matrix $P$ of $\Omega
_{\beta}$, we sample a vector $\mathbf{v}:= P\bolds{\beta}$, where $\mathbf
{v} = (L_{p}')^{-1}((L_{p}^{-1})P\bolds{\xi}_{\beta} + \mathbf{Z})$ with
$L_p$ a lower triangular matrix resulting from the Cholesky
decomposition of $P\Omega_{\beta,}$ and $\mathbf{Z}$ a standard normal
distributed vector, that is, $\mathbf{Z}\sim\mathcal{N}_{d}(\mathbf
{0},I_d)$. The original parameter vector of interest $\bolds{\beta}$
can be obtained after permuting vector $\mathbf{v}$ again. %i.e. $\bolds{
\citet{rue2005} suggested finding a permutation such that the matrix
is banded. However, we found that in our case the AMD algorithm is
more efficient with regard to computing time. Using MATLAB built-in
functions, the Cholesky decomposition with an approximate minimum
degree ordering takes 8 seconds on a Linux-Cluster 8-way Opteron
with 32 cores, while the Cholesky decomposition based on a banded
matrix takes 15~seconds. The traditional method without permutation
requires 118.5 seconds.
%Moreover, the AMD algorithm is a build-in function in the Cholesky
%decomposition function in MATLAB. These functions are efficiently
%parallelized and reduced the computational time enormously, especially
%in a high-dimensional problem.

%s4 #&#
\section{Simulation study}\label{sec4}
%s4.1 #&#
\subsection{Simulation setups}\label{sec4.1}
In this section we examine the performance of our approach for model 1.
We want to investigate whether the method\vadjust{\goodbreak} works correctly under the
correct model assumptions and how much influence the prior has on the
posterior estimation. We consider five different prior neighborhood
structures of $\bolds{\beta}_{\mathrm{usa}}$:\vspace*{9pt}
%The following five different prior structures for the variance $Q^{-1}(
%performance of the method:

(0) Independent model of $\bolds{\beta}_{\mathrm{usa}}$, $\psi
=0$ fixed, that is, $\bolds{\beta}_{\mathrm{usa}}\sim\mathcal
{N}_{d_{\mathrm{usa}}}(\bolds{\beta}_0,\break \frac{1}{\eta_{\mathrm
{usa}}}I_{d_{\mathrm{usa}}})$,

(1) Spherical neighborhood structure with reciprocal weight
function,

(2) Ellipsoidal neighborhood structure with reciprocal weight
function,

(3) Spherical neighborhood structure with exponential weight
function,

(4) Ellipsoidal neighborhood structure with exponential weight
function.\vspace*{9pt}

Note that the independent model of $\bolds{\beta}_{\mathrm{usa}}$
corresponds to the Bayesian ridge estimator as described in Section
\ref{sec3.3}. For the weight functions in (\ref{wfuncs}), we set
$D_x=D_y=300$ km and $D_z=150$ km for modeling ellipsoidal
neighborhood structures, and $D=150$ km for the spherical
neighborhood distance.\vspace*{9pt}

\textit{Setup} I: Assume the solution by
\citet{siglochmcquarrienolet2008}, denoted as
$\hat{\bolds{\beta}}{}^{\mathrm{LSQR}}_{\mathrm{usa}}$,
represents true mantle structure beneath North America. We use the
forward model $X
\hat{\bolds{\beta}}{}^{\mathrm{LSQR}}_{\mathrm{usa}}$ to
compute noise-free, synthetic data. Then, we generate two types of
noisy data, that is, $\mathbf{Y} =
\mathbf{X}\hat{\bolds{\beta}}{}^{\mathrm{LSQR}}_{\mathrm{usa}} +
\bolds{\varepsilon}$ with:\vspace*{9pt}

(A) Gaussian noise ($\bolds{\varepsilon} \sim\mathcal
{N}_{N}(\mathbf{0},\frac{1}{\phi_{\mathrm{tr}}}I_{N})$, $\phi_{\mathrm{tr}}=0.4$),
%(B) Residuals $\varepsilon_i$ from tomographic inversion, $\phi_{\mathrm{tr}},

(B) $t$-noise ($\bolds{\varepsilon}\sim t_{N}(\mathbf{0},I_N,\nu
_{\mathrm{tr}})$, $\nu_{\mathrm{tr}}=3$, corresponds to
$\phi_{\mathrm{tr}}=0.333$).\vspace*{9pt}

Although we add $t$-noise to our synthetic earth model
$\hat{\bolds{\beta}}{}^{\mathrm{LSQR}}_{\mathrm{usa}}$,
our posterior calculation is based on Gaussian errors. Additionally, we
compare two priors for $\bolds{\beta}_{\mathrm{usa}}\sim\mathcal
{N}_{d_{\mathrm{usa}}}(\bolds{\beta}_0,\frac{1}{\eta_{\mathrm
{usa}}}Q^{-1}(\psi))$ to examine the sensitivity of the posterior
estimates to the prior choices:\vspace*{9pt}

(a) $\bolds{\beta}_{0} \sim\mathcal{N}_{d_{\mathrm
{usa}}}(\hat{\bolds{\beta}}{}^{\mathrm{LSQR}}_{\mathrm{usa}}, 0.32^2
I_d)$,
%This prior mean allows for 32\% more deviations from the true
%velocities variation $\bolds{\beta}_{\mathrm{usa}}^{LSQR}$.

(b) $\bolds{\beta}_{0} = \mathbf{0}$ (spherically symmetric
reference model).\vspace*{9pt}
%The prior does not allow for much deviation from the reference Earth
%model.

The priors for the hyperparameters are set as follows:
$\psi\sim\mathcal{N}(10,0.2^2)$, $\phi\sim\Ga(1,0.1)$ resulting in
expectation and standard deviation of 10,
$\eta_{\mathrm{usa}}\sim\Ga(10,2)$ resulting in expectation of
5 and standard deviation of 1.6.\vspace*{9pt}

\textit{Setup} II: In this case we examine the performance
under known prior neighborhood structures. We construct a synthetic
true mantle model with two types of known prior neighborhood
structures: $\bolds{\beta}_{\mathrm{usa},\mathrm{tr}} \sim
\mathcal{N}_{d_{\mathrm{usa}}}(\hat{\bolds{\beta}}{}^{\mathrm{LSQR}}_{\mathrm
{usa}},\break\frac{1}{\eta_{\mathrm{usa},\mathrm{tr}}}Q^{-1}
(\psi_{\mathrm{tr}}))$
with $\eta_{\mathrm{usa},\mathrm{tr}}=0.18$ and $\psi_{\mathrm
{tr}}=10$ using:\vspace*{9pt}

(a) a spherical neighborhood structure for
$\bolds{\beta}_{\mathrm{usa},\mathrm{tr}}$ with reciprocal weights,

(b) an ellipsoidal neighborhood structure for $\bolds{\beta
}_{\mathrm{usa},\mathrm{tr}}$ with reciprocal weights.%\vspace*{9pt}

Again, Gaussian noise is added to the forward model, that is, $\mathbf{Y}
=\break X\hat{\bolds{\beta}}{}^{\mathrm{LSQR}}_{\mathrm{usa}} +
\bolds{\varepsilon}$, $\bolds{\varepsilon} \sim
\mathcal{N}_{N}(0,\frac{1}{\phi_{\mathrm{tr}}}I_{N})$,
$\phi_{\mathrm{tr}} = 0.4$. Posterior estimation is carried out
assuming the five different prior structures.

The number of MCMC iterations for scenarios in setups I and II
is 3000, thinning is 15, and burn-in after thinning is 100. For
convergence diagnostics we compute the trace, autocorrelation
and estimated density plots as well as the effective sample size (ESS)
using \texttt{coda} package in R for those samples. According to
\citet{brooksetal2011}, the ESS is defined by
$\mathrm{ESS}:=\frac{n}{1+2\sum_{k=1}^{\infty}\rho_k}$,\vspace*{1pt} with the original
sample size $n$ and autocorrelation $\rho_k<0.05$ at lag $k$. The
infinite sum can be truncated at lag $k$ when $\rho_k$
becomes smaller than 0.05 [\citet{kassetal1998,liu2008}].

\subsection{Performance evaluation measures}\label{sec4.2}
To evaluate the results, we use the standardized Euclidean norm for
both data and model misfits, $\|\cdot\|_{\Sigma_y}$ and
$\|\cdot\|_{\Sigma_{\beta}}$, respectively. The function
$\|\mathbf{x}\|_{\Sigma}$ of a vector $\mathbf{x}$ of mean $\bolds{\mu}$ and
covariance $\Sigma$ is called the \textit{Mahalanobis distance},
defined by $\|\mathbf{x}\|_{\Sigma}:=\break
\sqrt{(\mathbf{x}-\bolds{\mu})'\Sigma^{-1}(\mathbf{x}-\bolds{\mu})}$.
To include model complexity, we calculate the deviance information
criterion (DIC) [\citet{spiegelhalter2002}]. Let $\bolds{\theta}$
denote the
parameter vector to be estimated. Furthermore, the likelihood of the
model is denoted by $\ell(\mathbf{y}\mid\bar{\bolds{\theta}})$, where
$\bar{\bolds{\theta}}$ is the estimated posterior mean of
$\bolds{\theta}$, estimated by $\frac{1}{R}\sum_{r =
1}^{R}\bolds{\theta}^r$ with $R$ number of independent MCMC samples.
According to \citet{spiegelhalter2002} and \citet{celeux2006}, the
deviance is defined as $D(\bolds{\theta}) =
-2\log(\ell(\mathbf{y}\mid\bar{\bolds{\theta}})) + 2\log h(\mathbf{y})$. The
term $h(\mathbf{y})$ is a standardizing term which is a function of the
data alone and does not need to be known. Thus, for model comparison
we take $D(\bolds{\theta}) =
-2\log(\ell(\mathbf{y}\mid\bar{\bolds{\theta}}))$.
The effective number of parameters in the model, denoted by $p_D$, is
defined by
$p_D:= E_{\theta}[D(\bolds{\theta})] - D(\bar{\bolds{\theta}})$. The
term $E_{\theta}[D(\bolds{\theta})]$ is the posterior mean deviance
and is estimated by $\frac{1}{R}\sum_{r = 1}^{R}D(\bolds{\theta}^r)$.
This term can be regarded as a Bayesian measure of fit. In summary,
the DIC is defined as $\mathrm{DIC} = E_{\theta}[D(\bolds{\theta})] + p_D =
D(\bar{\bolds{\theta}}) + 2p_D$. The model with the smallest DIC is
the preferred model under the trade-off of model fit and model
complexity.

%
%t1 #&#
\begin{sidewaystable}
\textwidth=\textheight
\tablewidth=\textwidth
\caption{Posterior estimation results of the simulation study under
setups \textup{I} and \textup{II}, using synthetic earth models.
The posterior mode of the velocity parameters is denoted as $\hat{\bolds
{\beta}}$. The quantities $\hat{\bolds{\beta}}_L$ and $\hat{\bolds
{\beta}}_U$
are lower and upper quantiles of the $90\%$ credible interval of the
MCMC estimates, respectively}\label{chap3case1}
\begin{tabular*}{\tablewidth}{@{\extracolsep{\fill}}lccccd{3.2}cc@{}}
\hline
\multicolumn{8}{@{}c@{}}{\textbf{Setup I}}\\
\hline
& \textbf{Prior} & & & & & & \textbf{Mode}\\
\textbf{Noises} & \textbf{struct} & $\bolds{\|\mathbf{y} - X\hat
{\bolds{\beta}}\|_{\Sigma_y}}$ & $\bolds{\|\mathbf{y} - X\hat
{\bolds{\beta}}_{L}\|_{\Sigma_y}}$ & $\bolds{\|\mathbf{y} - X\hat
{\bolds{\beta}}_{U}\|_{\Sigma_y}}$ & \multicolumn{1}{c}{$\bolds{\|\hat{\bolds
{\beta}}-\bolds{\beta}_{\mathrm{tr}}\|_{\Sigma_{\beta}}}$} &
\textbf{DIC} & $\bolds{\hat{\eta}_{\mathrm{usa}}}$ \\
\hline
\multicolumn{8}{@{}c@{}}{(a) $\bolds{\beta}\sim\mathcal
{N}_d(\bolds{\beta}_{0},\frac{1}{\eta_{\mathrm{usa}}}Q^{-1}(\psi))$, $\bolds
{\beta}_{0} \sim\mathcal{N}_d(\hat{\bolds{\beta}}{}^{\mathrm{LSQR}}_{\mathrm{usa}},
0.32^2 I_d)$}\\
[6pt]
(A) Gaussian noise & (0) &232.28&312.82&308.27&91.30 &103,748&9.28 \\
\quad$\varepsilon_i \sim N(0,1/\phi_{\mathrm{tr}})$ & (1) &231.71&258.32&255.74&349.70&103,467&2.89 \\
\quad$\phi_{\mathrm{tr}}=0.4$, $\eta_{\mathrm{usa}}$, & (2)
&231.67&264.81&261.59&200.34&103,442&0.11 \\
\quad$\psi$ unknown & (3) &231.77&263.89&260.96&256.52&103,478&2.70 \\
& (4)
&231.70&267.81&264.44&185.04&103,456&0.20 \\
[4pt]
(B) $t$-noises & (0) &227.74&602.35&604.21&46.83 &112,436&0.57 \\
\quad$\varepsilon_i\sim t(0,1,\nu_{\mathrm{tr}})$ & (1) &228.58&443.00&443.02&69.50
&112,226&0.09 \\
\quad$\phi_{\mathrm{tr}}=0.33$, & (2) &228.57&437.58&437.80&57.83 &112,118&0.01 \\
\quad$\nu_{\mathrm{tr}}=3$, $\eta_{\mathrm{usa}}$, & (3) &228.51&450.67&450.27&62.03 &112,175&0.11 \\
\quad$\psi$ unknown & (4) &228.52&445.22&445.42&55.87
&112,126&0.01 \\
[6pt]
\multicolumn{8}{@{}c@{}}{(b) $\bolds{\beta}\sim\mathcal
{N}_d(\mathbf{0},\frac{1}{\eta_{\mathrm{usa}}}Q^{-1}(\psi))$}\\
[6pt]
%{\bolds{\beta}}\|_{\Sigma_y}$} & \multirow{2}*{$\|\mathbf{y} - X\hat
%{\bolds{\beta}}_{L}\|_{\Sigma_y}$} &\multirow{2}*{$\|\mathbf{y} - X\hat
%{\bolds{\beta}}_{U}\|_{\Sigma_y}$} &\multirow{2}*{$\|\hat{\bolds
%{\beta}}-\bolds{\beta}_{\mathrm{tr}}\|_{\Sigma_{\beta}}$} &\multirow
%{2}*{DIC} &mode \\
%& struct & & & & & &$\hat{\eta}_{\mathrm{usa}}$ \\ \hline
(A) Gaussian noise & (0) &234.33&635.99&632.19&50.53&106,563&0.59 \\
\quad$\varepsilon_i \sim N(0,1/\phi_{\mathrm{tr}})$ & (1) &233.46&458.55&454.42&40.53&105,365&0.09 \\
\quad$\phi_{\mathrm{tr}}=0.4$, $\eta_{\mathrm{usa}}$, & (2)
&233.36&449.35&444.06&42.45&105,200&0.01 \\
\quad$\psi$ unknown & (3) &233.52&466.36&462.04&38.55&105,357&0.12 \\
& (4)
&233.40&458.05&452.60&42.71&105,256&0.01 \\
[4pt]
(B) $t$-noises & (0) &226.53&831.85&832.84&40.10&113,023&0.19 \\
\quad$\varepsilon_i\sim t(0,1/\phi_{\mathrm{tr}},\nu_{\mathrm{tr}})$ & (1)
&227.60&599.53&598.99&33.50&112,575&0.03 \\
\quad$\phi_{\mathrm{tr}}=0.33$, & (2) &227.56&596.04&595.78&33.62&112,512&0.00 \\
\quad$\nu_{\mathrm{tr}}=3$, $\eta_{\mathrm{usa}}$, & (3) &227.61&606.60&605.77&33.48&112,541&0.03 \\
\quad$\psi$ unknown & (4)
&227.55&607.03&606.69&34.03&112,536&0.00 \\
\hline
\end{tabular*}
\end{sidewaystable}

\setcounter{table}{0}
\begin{sidewaystable}
\textwidth=\textheight
\tablewidth=\textwidth
\caption{(Continued)}
\begin{tabular*}{\tablewidth}{@{\extracolsep{\fill}}lccccccd{2.2}@{}}
\hline
\multicolumn{8}{@{}c@{}}{\textbf{Setup II}}\\
\hline
& \textbf{Prior} & & & & & & \multicolumn{1}{c@{}}{\textbf{Mode}}\\
\textbf{Noises} & \textbf{struct} & $\bolds{\|\mathbf{y} - X\hat
{\bolds{\beta}}\|_{\Sigma_y}}$ & $\bolds{\|\mathbf{y} - X\hat
{\bolds{\beta}}_{L}\|_{\Sigma_y}}$ & $\bolds{\|\mathbf{y} - X\hat
{\bolds{\beta}}_{U}\|_{\Sigma_y}}$ & $\bolds{\|\hat{\bolds
{\beta}}-\bolds{\beta}_{\mathrm{tr}}\|_{\Sigma_{\beta}}}$ &
\textbf{DIC} & \multicolumn{1}{c@{}}{$\bolds{\hat{\eta}_{\mathrm{usa}}}$} \\
\hline
%{\bolds{\beta}}\|_{\Sigma_y}$} &\multirow{2}*{$\|\mathbf{y} - X\hat
%{\bolds{\beta}}_L\|_{\Sigma_y}$}&\multirow{2}*{$\|\mathbf{y} - X\hat
%{\bolds{\beta}}_U\|_{\Sigma_y}$}&\multirow{2}*{$\|\hat{\bolds{\beta
%}}-\bolds{\beta}_{\mathrm{tr}}\|_{\Sigma_{\beta}}$} &\multirow{2}*{DIC} &mode
%& struct & & & & & &$\hat{\eta}_{\mathrm{usa}}$ \\ \hline
%
\multicolumn{8}{@{}c@{}}{(a) $\bolds{\beta}\sim\mathcal
{N}_d(\bolds{\beta}_{0},\frac{1}{\eta_{\mathrm{usa}}}Q^{-1}(\psi))$ with a
spherical neighborhood structure for $Q$}\\
[6pt]
Gaussian noise & (0) &279.38&575.55&520.63&40.06&129,433&1.09\\
\quad$\varepsilon_i \sim N(0,1/\phi_{\mathrm{tr}})$ & (1)
&279.82&439.32&389.22&35.83&128,882&0.20 \\
\quad$\phi_{\mathrm{tr}}=0.4$, & (2) &279.78&475.57&422.49&36.95&129,034&0.01 \\
\quad$\eta_{\mathrm{usa}}=0.18$, & (3) &279.96&455.81&404.30&36.23&128,986&0.22 \\
\quad$\psi=10$ & (4) &279.79&481.30&427.90&37.10&129,066&0.02 \\
[6pt]
\multicolumn{8}{@{}c@{}}{(b) $\bolds{\beta}\sim\mathcal
{N}_d(\bolds{\beta}_{0},\frac{1}{\eta_{\mathrm{usa}}}Q^{-1}(\psi))$ with an
ellipsoidal neighborhood structure for $Q$}\\
[6pt]
%{\bolds{\beta}}\|_{\Sigma_y}$} &\multirow{2}*{$\|\mathbf{y} - X\hat
%{\bolds{\beta}}_L\|_{\Sigma_y}$}&\multirow{2}*{$\|\mathbf{y} - X\hat
%{\bolds{\beta}}_U\|_{\Sigma_y}$}&\multirow{2}*{$\|\hat{\bolds{\beta
%}}-\bolds{\beta}_{\mathrm{tr}}\|_{\Sigma_{\beta}}$} &\multirow{2}*{DIC} &mode
%& struct & & & & & &$\hat{\eta}_{\mathrm{usa}}$ \\ \hline
Gaussian noise & (0) &234.71&305.10&292.47&27.71&104,662&11.84\\
\quad$\varepsilon_i \sim N(0,1/\phi_{\mathrm{tr}})$ & (1)
&234.37&257.34&249.46&26.10&104,262&4.19 \\
\quad$\phi_{\mathrm{tr}}=0.4$, & (2) &234.27&251.55&244.20&24.59&104,152&0.30 \\
\quad$\eta_{\mathrm{usa}}=0.18$, & (3) &234.41&260.03&251.59&25.72&104,284&4.22 \\
\quad$\psi=10$ & (4) &234.30&253.50&245.76&24.50&104,173&0.53 \\
\hline
\end{tabular*}
\end{sidewaystable}

\subsection{Results and interpretations}\label{sec4.3}
The first two blocks in Table \ref{chap3case1} illustrate posterior
estimation results for setup I. It shows that the estimation method
with ellipsoidal prior structures (2) and (4) turn out to be the
most adequate, according to the DIC criterion. The standardized data
misfit criteria $\|\cdot\|_{\Sigma_y}$ given the estimated posterior
mode $\hat{\bolds{\beta}}$ show similar results in all scenarios.
However, this measure ignores the uncertainty of
$\bolds{\beta}_{\mathrm{usa}}$. The criteria
$\|\mathbf{y}-X\hat{\bolds{\beta}}_{L}\|_{\Sigma_y}$ and
$\|\mathbf{y}-X\hat{\bolds{\beta}}_{U}\|_{\Sigma_y}$ show the data misfit
given the $90\%$ credible interval with lower and upper quantile
posterior estimates $\hat{\bolds{\beta}}_{L}$ and
$\hat{\bolds{\beta}}_{U}$, respectively. These estimates give a range
of the data misfit for all possible posterior solutions of
$\bolds{\beta}_{\mathrm{usa}}$ and show that methods with
independent prior generally yield larger ranges of misfit values
than the ones with spatial structures. This indicates that the
credible intervals of methods with spatial priors can fit the data
better.
Further, methods with spatial priors in setup~I(b) show smaller
model misfit under $\|\cdot\|_{\Sigma_{\beta}}$ than ones with
independent prior, while in setup I(a) results with independent
priors are better. Generally, estimated posterior modes of
$\eta_{\mathrm{usa}}$ vary considerably due to the different
prior assumptions. Models with ellipsoidal neighborhood structures
have a stronger prior (in the sense of a smaller prior variance)
than models with spherical neighborhood structure. Similarly, models
with reciprocal weights have a stronger regularization toward the
prior mean than models with exponential weights. This means that the
posterior estimates of $\eta_{\mathrm{usa}}$ adapt to different
prior settings. Moreover, we notice that the estimate of the spatial
dependence parameter $\psi$ depends strongly on its prior, as the
prior mean is close to the posterior estimates of $\psi$ in all
scenarios. The last two blocks in Table \ref{chap3case1} illustrate
results from setup II assuming known spatial structure including
hyperparameters. The DIC values indicate that our approach
correctly detects the underlying prior structures [in (a) it is
prior structure (1), in (b) it is prior structure (2)]. We can also
observe that our approach estimates the hyperparameters correctly.
Estimated posterior modes of the parameters from the identified
model are close to their true values.\looseness=-1

% \begin{minipage}{0.2\linewidth}
% \includegraphics[scale=0.6]{./graphics/paper/case1/true_model.pdf}
% \end{minipage}
% \vline
% \begin{minipage}{0.39\linewidth}
%
%
% \end{minipage}
% \vline
% \begin{minipage}{0.39\linewidth}
%
%
% \end{minipage}
% \caption{Posterior mode of velocity deviation from the reference
%Earth model (in $\%$),
% estimated using ellipsoidal prior structure with reciprocal weights.
% Left column: true solution. Middle columns: tomographic results of
%Setup I (a) using informative prior mean;
% Right columns: tomographic results of Setup I (b) assuming prior mean
%$\bolds{\beta}_0 = \mathbf{0}$.}\label{chap3pic1}

% \begin{minipage}{1.0\linewidth}
%
%
% \end{minipage}
% \caption{Posterior mode of tomographic solution estimated using
%ellipsoidal prior structure with reciprocal weights
% and informative prior mean}\label{chap3pic1}
% \begin{minipage}{1.0\linewidth}
%
%
% \end{minipage}
% \caption{Posterior mode of tomographic solution estimated using
%ellipsoidal prior structure with reciprocal weights
% and reference Earth model as prior mean}\label{chap3pic1beta0}

%prior of $\bolds{\beta}$ independent model $\bolds{\beta}\sim\mathcal{N}_d(
%$\bolds{\beta}_0=\bolds{\beta}^{LSQR}_{usa}+\frac{1}{\sqrt{10}}\bolds{
%quantiles and regions
%significantly differ from the reference model $\bolds{\beta} = \mathbf{0}$
%for depth 200km, 400km and 600km.}\label{caseI0}
% \includegraphics[width=8cm]{./graphics/paper/case1/plots_1_0.pdf}
% \vline
%
%
%%\caption{CASE I (0): }\label{caseI0qnt}
%
%
% \vline
%
%

Generally, tomographic images illustrate velocity parameters as
deviation of the solution from the spherically symmetric reference
model (in \%). Blue colors represent zones that have faster seismic
velocities than the reference earth model, while red colors denote
slower velocities. Physically, blue colors usually imply that those
regions are colder than the default expectation for the
corresponding mantle depth, while red regions are hotter than
expected. In our simulation study, we assumed the true earth to be
represented by the solution of \citet{siglochmcquarrienolet2008},
shown in the left column of Figure \ref{chap3pic1}. The middle and
right columns of Figure \ref{chap3pic1} illustrate the estimated
posterior modes $\hat{\bolds{\beta}}_{\mathrm{usa}}$ from setup I
with ellipsoidal neighborhood structure and reciprocal weight for
both Gaussian and $t$-noises, respectively. They show that the
parameter estimates from Gaussian noises are close to the true
solution, while the solution from the $t$-noises tends to
overestimate the parameters. The magnitude of mantle anomalies is
overestimated but major structures are correctly recovered. The same
effect can be seen in the last column of Figure \ref{chap3pic1}
which displays the estimated posterior modes of the tomographic
solutions in setup I(b). We have overestimation since the noise is
not adequate to the Gaussian model assumption. Moreover, we also
observe that tomographic solutions with the prior mean
$\bolds{\beta}_0 \neq\mathbf{0}$ are smoother than the ones with the
prior mean $\bolds{\beta}_0 = \mathbf{0}$.

%
%f4 #&#
\begin{figure}

\includegraphics{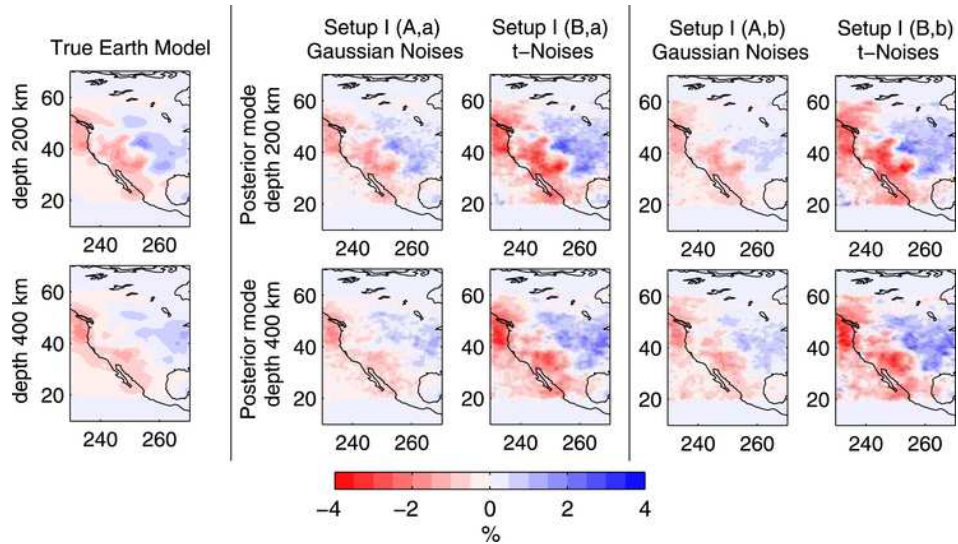}

\caption{Mantle models resulting from the simulation study. Left
column shows the ``true'' model, used to generate the synthetic data.
The unit on the color bar is velocity deviation $\bolds{\beta}$ in \%
from the spherically symmetric reference model.
All other columns show the posterior mode of velocity deviation $\bolds
{\beta}$, estimated using ellipsoidal prior structure with reciprocal
weights. Middle columns show results for setup \textup{I(a)}, which uses the
prior mean $\bolds{\beta}\neq\mathbf{0}$. Right columns show results for
setup~\textup{I(b)} assuming prior mean $\bolds{\beta}_0 = \mathbf{0}$.}
\label{chap3pic1}
\end{figure}
%

%
%f5 #&#
\begin{figure}

\includegraphics{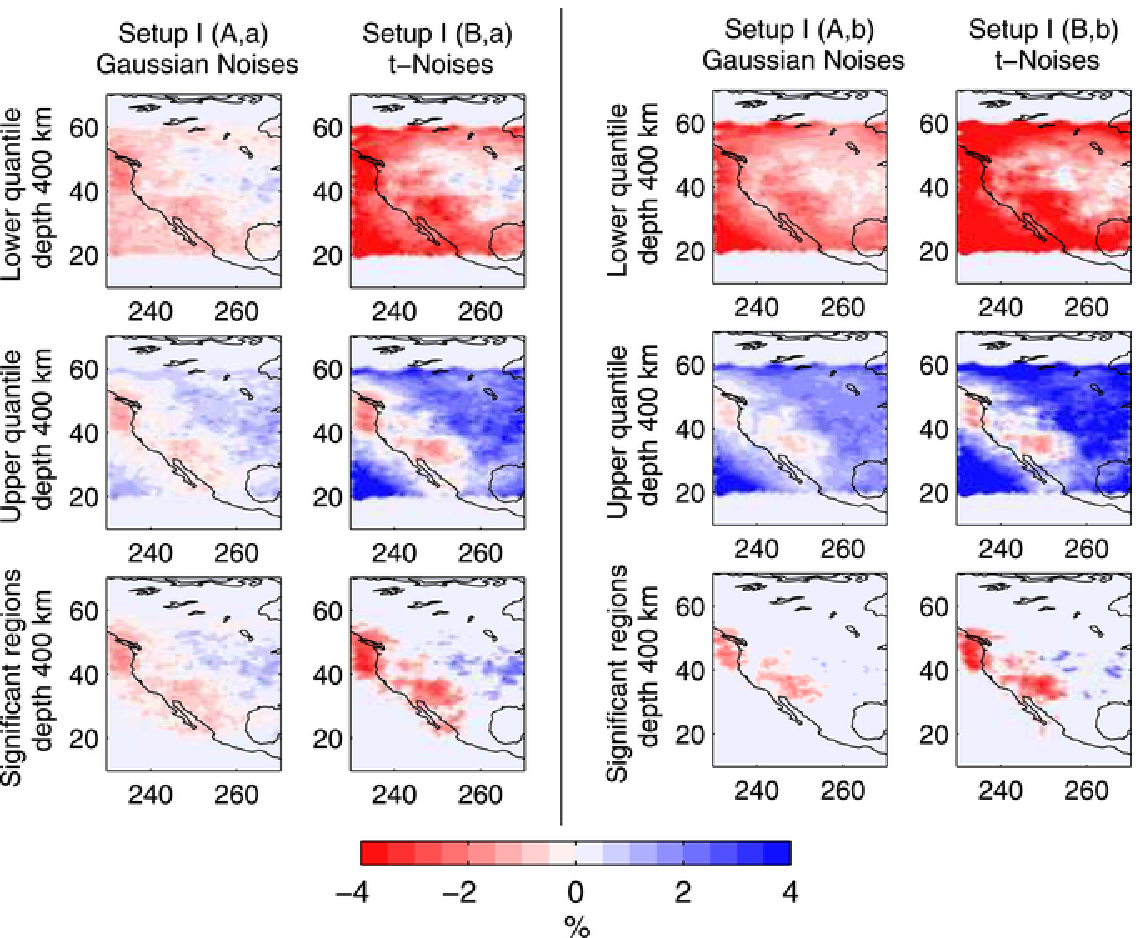}

\caption{Continuation of Figure \protect\ref{chap3pic1}. The maps show
velocity deviation in \% from the reference Earth model.
Left half shows the results under setup \textup{I(a)}, which uses the prior
mean $\bolds{\beta}_0 \neq\mathbf{0}$; right half describes setup \textup{I(b)},
which uses prior mean $\bolds{\beta}_0 =\mathbf{0}$. First and second
rows map out the lower and upper quantiles of the 90\% confidence
interval. Third row shows the posterior mode of velocity structure
$\bolds{\beta}$, but rendered only in regions that differ
significantly from the reference model, according to the 90\%
confidence interval.} \label{chap3quantiles}
\end{figure}

Figure \ref{chap3quantiles} shows estimated credible intervals for
the solutions of Figure~\ref{chap3pic1}. Credible intervals for
solutions with $t$-noises are larger than those for the Gaussian
noises, as indicated by the darker shades of blue/red colors, which
denote higher/lower quantile estimates. This implies that parameter
uncertainty is greater if noise does not fit the model assumption.
The same effect can be seen for results with the prior mean
$\bolds{\beta}_0 = \mathbf{0}$. The bottom row of Figure
\ref{chap3quantiles} maps out how the regions differ from the reference
model with $90\%$ posterior probability. For model-conform Gaussian
distributed noises and informative prior mean, more regions differ
from the reference model with $90\%$ posterior probability than if
we added $t$-noise or used the less informative prior. In the case of an
informative prior and/or correctly modeled noise, we achieve more
certainty about the velocity deviations from the reference earth
model.

%s5 #&#
\section{Application to real seismic travel time data}\label{sec5}
% What are USArray, how the data are preprocessed (data selection)
% Who has used those data, what are their methods and what are the
%results
% Comparison of our results and the existing tomographic solution

In this section we apply our MCMC approach to actually measured
travel time data.

The measurements are a subset of those generated by
\citet{siglochmcquarrienolet2008}. We use the same wave paths, but
only measurements made on the broadband waveforms, whereas they
further bandpassed the data for finite-frequency measurements and
also included amplitude data [\citet{siglochnolet2006}]. Most
stations are located in the western U.S., as part of the
largest-ever seismological experiment (USArray), which is still in
the process of rolling across\vadjust{\goodbreak} the continent from west to east.
Numerous tomographic studies have incorporated USArray data---the
ones most similar to ours are \citet{burdicketal2008},
\citet{siglochmcquarrienolet2008}, \citet{tiansiglochnolet2009},
and \citet{schmandt2010}. All prior studies obtained their solutions
through least-squares minimization, which yields no uncertainty
estimates. Here we use 53,270 broadband travel time observations to
estimate velocity structure under western North America (over 11,000
parameters), plus source corrections for 529 events (2116~parameters). We conduct our Bayesian inversion following two
different scenarios:
%in Model 1 we only estimate the earth structure parameters $\bolds{
%In Model 2 we estimate the earth structure and source correction
%parameters jointly.
%Our approach is as well under Gaussian assumption. The prior
%specifications are as follows:
%

\textit{Model} 1: We only invert for earth structural
parameters. For the velocity parameters we assume
$\bolds{\beta}\sim\mathcal{N}_{d_{\mathrm{usa}}}(\hat{\bolds{\beta
}}{}^{\mathrm{LSQR}}_{\mathrm{usa}},\frac{1}{\eta_{\mathrm
{usa}}}Q^{-1}(\psi))$
as in (\ref{stat1}) with
$\psi\sim\mathcal{N}(10,\allowbreak0.5^2)\mathbbm{1}(\psi>0)$, $\phi\sim
\Ga(1,0.1)$ and $\eta_{\mathrm{usa}}\sim\Ga(10,2)$.

\textit{Model} 2: We invert\vspace*{1pt} for both earth structural
parameters and the source corrections. The prior distributions are
set to
$\bolds{\beta}\sim\mathcal{N}_d(\hat{\bolds{\beta}}{}^{\mathrm
{LSQR}},\Sigma_{\beta})$
as in (\ref{stat2}) and $\Sigma_{\beta}$ as defined in
(\ref{sigmabeta}). For $\psi$, $\phi$ and
$\eta_{\mathrm{usa}}$, we adopt the same distribution as in
model 1. For the parameters of the source corrections we adopt
$\eta_{\mathrm{hyp}}\sim\Ga(1,5)$ and
$\eta_{\mathrm{time}}\sim\Ga(10,2)$.

%f6 #&#
\begin{figure}

\includegraphics{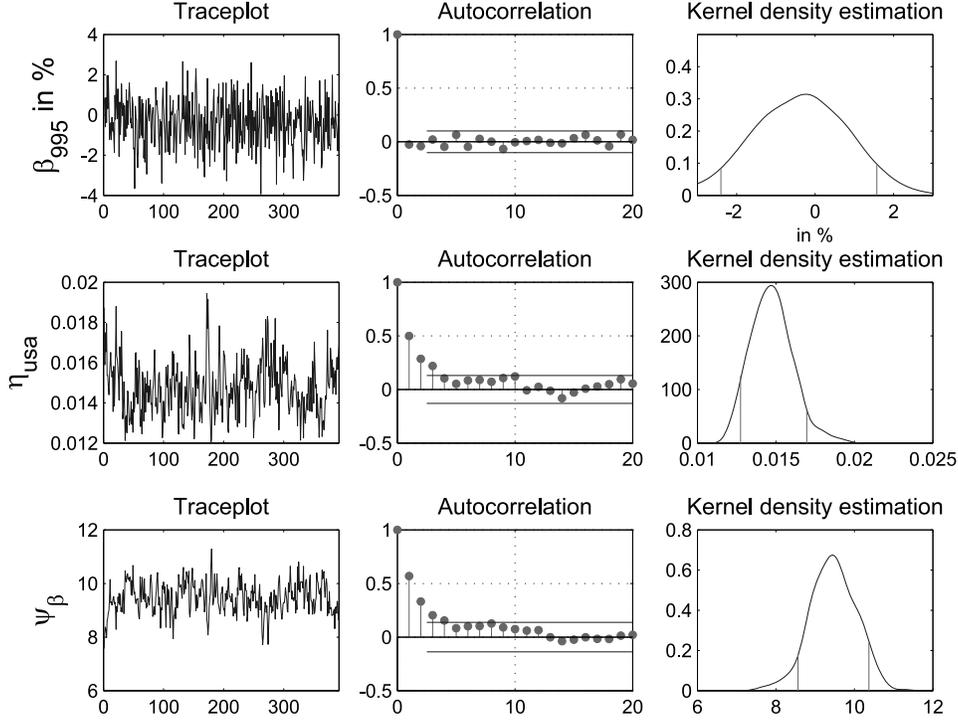}

\caption{Convergence diagnostics: trace plot, autocorrelation and
kernel density estimation of the parameters $\bolds{\beta}_{\mathrm
{usa}}$ at node $955$, $\eta_{\mathrm{usa}}$ and $\psi_{\beta}$.
For 10,000 MCMC iterations the samples shown in plots are based on a
burn-in of 200 and a thinning rate of 25.}
\label{convdiag}
\end{figure}

We use the same five prior structures (0)--(4) as in the simulation
study and run the MCMC algorithm for 10,000 iterations.
The high-dimensional $\bolds{\beta}$ vector can be sampled efficiently
in terms of ESS with low burn-in and thinning rates thanks to the
efficient Gibbs sampling scheme in (\ref{betafullcond}). However,
the hyperparameters, for example, $\psi_{\beta}$, are more difficult
to sample.
To achieve a good mixing, we applied a burn-in of 200 and a
thinning rate of 25 (393 samples for each parameter) in our analysis.
On average, the effective sample size ESS values for $\bolds{\beta
}_{\mathrm{usa}}$,
$\bolds{\beta}_{\mathrm{hyp}}$ and
$\bolds{\beta}_{\mathrm{time}}$ are about $393$, $393$ and $327$,
respectively, which indicate very low autocorrelations for most of
the parameters. The ESS of both $\eta_{\mathrm{usa}}$ and
$\psi_{\beta}$ is about 103, while both $\eta_{\mathrm{hyp}}$
and $\phi$ have good mixing characteristics with ESS values equal to
the sample size, and $\eta_{\mathrm{time}}$ has ESS value
equal to 165. Figure \ref{convdiag} shows as examples the
parameters $\beta_{\mathrm{usa},955}$ at node $955$,
$\eta_{\mathrm{usa}}$ and $\psi_{\beta}$. The computing cost of
our algorithm is about $O(n^4)$. Sampling model 1 with about 9000
parameters, our algorithm needs 12 hours in 10,000 runs on a 32-core
cluster, while under the same condition it needs 38 hours for model
2.

%t2 #&#
\begin{table}
\tabcolsep=0pt
\caption{Posterior estimation results for the inversion using real
data, under models 1 and 2 specifications}\label{chap41}
\begin{tabular*}{\tablewidth}{@{\extracolsep{\fill}}lccccccc@{}}
\hline
%%\multirow{2}*{Sc\#} & \multirow{2}*{$\nu_1$} & \multirow{2}*{$\nu_2$}
%& \multirow{2}*{$\nu$} & \multirow{2}*{$\rho$} & \multirow{2}*{Method}
%multicolumn{4}{c|}{$\widehat{mad}$} & \multicolumn{4}{c|}{$
%% & & & & & & $\hat{\nu}_1$ & $\hat{\nu}_2$ & $\hat{\nu}$ & $\hat{
\multicolumn{8}{@{}c@{}}{\textbf{Model 1}}\\
\hline
\textbf{Prior} & & & & & \textbf{Mode} &\textbf{Mode} &\textbf{Mode}\\
\textbf{struct} & $\bolds{\|\mathbf{y} - X\hat{\bolds{\beta}}\|_{\Sigma
_y}}$ & $\bolds{\|\mathbf{y} - X\hat{\bolds{\beta}}_L\|_{\Sigma
_y}}$ & $\bolds{\|\mathbf{y} - X\hat{\bolds{\beta}}_U\|_{\Sigma
_y}}$ & \textbf{DIC} & $\bolds{\hat{\phi}}$ &$\bolds{\hat{\eta}_{\mathrm{usa}}}$
&$\bolds{\hat{\psi}}$ \\
\hline
(0) &228.46&490.92&490.46&102,928&0.40&1.40&$-$\\
(1) &229.14&389.73&390.21&104,096&0.39&0.20&9.63 \\
(2) &228.72&464.73&465.67&103,466&0.40&0.01&9.98 \\
(3) &228.90&430.78&431.34&103,749&0.39&0.17&9.63 \\
(4) &228.74&471.50&472.37&103,408&0.40&0.01&9.98 \\
\end{tabular*}
\begin{tabular*}{\tablewidth}{@{\extracolsep{\fill}}lccccccd{2.2}cc@{}}
\hline
\multicolumn{10}{@{}c@{}}{\textbf{Model 2} }\\
\hline
\textbf{Prior} &&&&& \textbf{Mode} & \textbf{Mode} & \multicolumn{1}{c}{\textbf{Mode}} &
\textbf{Mode} & \textbf{Mode}\\
\textbf{struct} & $\bolds{\|\mathbf{y} - X\hat{\bolds{\beta}}\|_{\Sigma
_y}}$ & $\bolds{\|\mathbf{y} - X\hat{\bolds{\beta}}_L\|_{\Sigma
_y}}$ & $\bolds{\|\mathbf{y} - X\hat{\bolds{\beta}}_U\|_{\Sigma
_y}}$ & \textbf{DIC} & $\bolds{\hat{\phi}}$ &
$\bolds{\hat{\eta}_{\mathrm{usa}}}$ &\multicolumn{1}{c}{$\bolds{\hat
{\psi}}$} &$\bolds{\hat{\eta}_{\mathrm{hyp}}}$ &$\bolds{\hat{\eta}_{\mathrm
{time}}}$ \\
\hline
(0) &225.40&483.96&488.35&93,788&0.49&1.15&\multicolumn{1}{c}{$-$} &0.01&5.01 \\
(1) &225.76&515.29&524.61&94,993&0.48&0.10&9.63 &0.01&4.53 \\
(2) &225.48&498.96&501.45&94,374&0.48&0.00&9.55 &0.01&4.70 \\
(3) &225.61&503.20&512.01&94,669&0.48&0.11&9.63 &0.01&4.53 \\
(4) &225.44&496.69&498.97&94,312&0.49&0.01&10.00&0.01&4.70 \\
\hline
\end{tabular*}
\end{table}

Table \ref{chap41} shows the results from model 1 (estimation of
earth structure) and model~2 (earth structure plus source
corrections). For both models, results from the independent prior
structures, corresponding to the Bayesian ridge estimator, provide
the best fit according to the DIC criterion. We also run the model 1
with prior mean $\bolds{\beta}_0=\mathbf{0}$ (the spherically symmetric
reference model) and different covariance structures (0)--(2). The
DIC results for priors (0), (1) and (2) are 103,100, 103,700 and
103,370, respectively. Two reasons may explain the selection of
prior~(0): (1)~the data has generally more correlation structure than
the i.i.d. Gaussian assumption, which can not be solely explained by
the spatial prior structure of the $\bolds{\beta}$-fields. However, in
our simulation study where different prior structures and the
corrected data error are applied (Table \ref{chap3case1}), the DIC was able to
identify the correct models; (2) Since the data are noisy, fitting
could be difficult without a shrinkage prior. The prior in (0)
can be compared to shrinkage in the ridge regression, which is the
limiting case of priors in (1) to (4). Priors in (1) to (4) do not
shrink the solutions of $\bolds{\beta}$-fields as much as prior (0).
They better reflect the uncertainty since the prior covariances in
(1)--(4) are larger than variances in prior (0) in regions
that have no data (no neighboring nodes), and smaller in regions
with lots of data (lots of neighboring nodes).

Furthermore, the standardized data misfit criteria
\mbox{$\|\cdot\|_{\Sigma_y}$} do not show much difference between models
with different prior specifications. According to the estimated
$90\%$ credible interval, estimates using spherical prior structure
show a smaller range of data misfit in model 1, whereas in model 2,
the independence prior shows a better result. Since our method
assumes i.i.d. Gaussian errors, the resulting residuals might not be
optimally fitted as expected. With regard to computational time, the
independent prior model has a definite advantage over other priors
in both models 1 and 2. The general advantage of our Bayesian
method is that the independent model yields an estimate given as the
ratio between the variance of the data and the variance of the
priors corresponding to ridge estimates with \textit{automatically
chosen shrinkage} described in Section \ref{sec3.3}, whereas in
\citet{aster2005}, \citet{siglochmcquarrienolet2008},
\citet{bodinetalJGR2012} and all other prior work, the shrinkage
parameter (strength of regularization) had to be chosen by the user
a priori.

%f7 #&#
\begin{figure}

\includegraphics{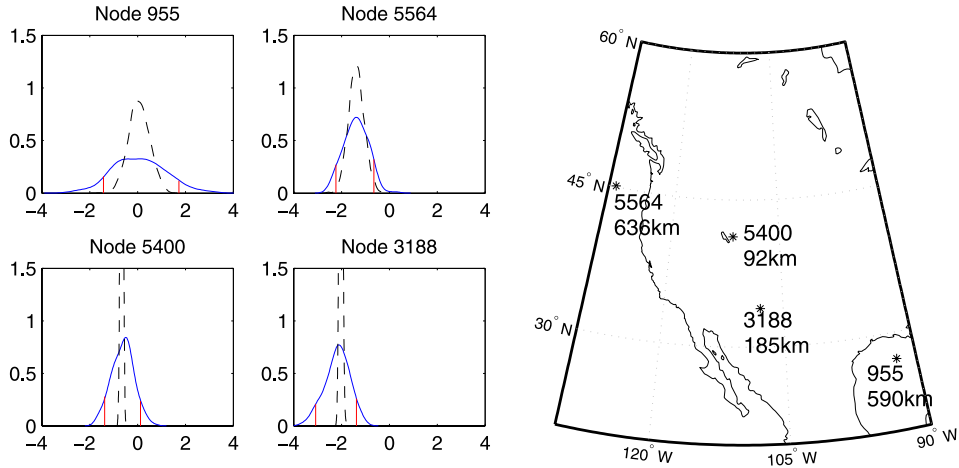}

\caption{Results of the Bayesian tomography using real travel time
observations. Left: estimated posterior density of $\bolds{\beta
}_{\mathrm{usa}}$ at a few selected model nodes, whose locations and
depths are indicated on the map. Unit on the x-axes is velocity
deviation in \%.
Dashed lines: prior density, the prior variance can be very small if
number of neighbors is large. Solid lines: posterior density with $90\%
$ credible intervals.}
\label{postmodepos}
\end{figure}

%f8 #&#
\begin{figure}

\includegraphics{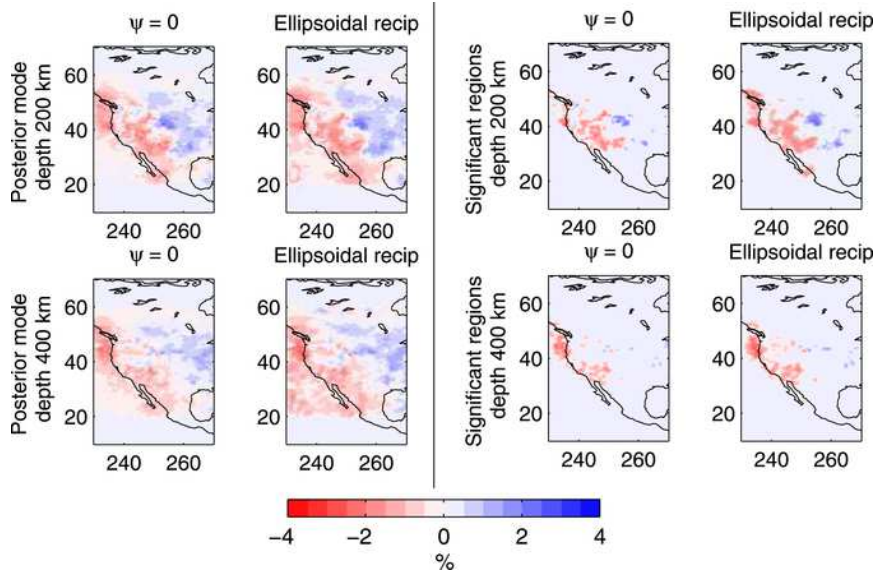}

\caption{Results of the Bayesian tomography using real travel time
data. All maps show estimated velocity deviation from the reference
earth model IASP91 (in $\%$).
Left columns: estimated posterior mode of velocity deviation, for the
scenario of model 2.
Right columns: same posterior mode, but rendered are only regions that
differ from the reference model with $90\%$ posterior probability.}
\label{chap4model2}
\end{figure}

Figure \ref{postmodepos} shows the estimated posterior and prior
densities of parameters in model 2, at four different locations of
varying depth. We see that parameters at locations with good ray
coverage, for example, node 5400 and node 3188, have smaller credible
intervals than parameters at locations with no ray coverage, for example,
node 5564 and node 995 beneath the uninstrumented oceans.
Geologically, the regions between node 5400 and node 3188 are well
known to represent the hot upper mantle, where seismic waves travel
slower than the reference velocity. This is consistent with our
results in Figure \ref{postmodepos}: the fact that
$\bolds{\beta}=\mathbf{0}$ does not fall inside the $90\%$ credible
intervals indicates a velocity deviation from the spherically
symmetric reference model with high posterior probability.
Figure \ref{postmodepos} shows that the posterior is more diffuse
than the prior. As mentioned in Section \ref{sec3.1}, the spatial prior for
$\bolds{\beta}$ depends on distance of neighboring nodes, number of
neighbors and orientation. The variance can be very small if the number
of neighbors is very
large, as shown in Figure \ref{weightsfunc}. Incorporating data, the
information about $\bolds{\beta}$ is updated
and thus may yield more diffuse posteriors than the priors, as we
see here. The left half of Figure \ref{chap4model2} shows the
estimated posterior modes of mantle structure obtained by model 2,
for independent and for ellipsoidal priors with reciprocal weights.
The right half of Figure \ref{chap4model2} extracts only those
regions that differ from the reference model according to the $90\%$
credible interval.
% Here we used the fact that $\bolds{\beta}_{usa}=\mathbf{0}\notin$ $90\%$
%credible intervals indicates velocity variation from the reference
%Earth model.
The ellipsoidal prior results in higher certainty of velocity
deviations at a depth of 200 km compared to the independence prior.
At a depth of 400 km, the credible regions resemble each other more
strongly. This confirms geological arguments that deeper regions of
the mantle are more homogeneous and do not differ as much from the
spherically symmetric reference model as shallower regions.

Many lines of geoscientific investigations provide independent
confirmation of the significantly anomalous regions of
Figure \ref{chap4model2}. The red areas map out the hot upper
mantle under the volcanic, extensional Basin and Range province and
Yellowstone; the blue anomalies map out the western edge of the old
and cool North American craton.

The overall comparison of our solutions to earlier least-squares
inversions, for example, the model by \citet{sigloch2008} shown in the
left column of Figure~\ref{chap3pic1}, confirms that Bayesian
inversion successfully retrieves the major features of mantle
structure. The images are similar, but the major advantage and
novelty of our approach is that it also quantifies uncertainties in
the solution (which we have chosen to visualize as credible
intervals here).

%s6 #&#
\section{Discussion and outlook}\label{sec6}
% Presents and future, extension of the model: what has been done and
%what has to be done.
Uncertainty quantification in underdetermined, large inverse
problems is important, since a single solution is not sufficient for
making conclusive judgements. Two central difficulties for MCMC
methods have always been the dimensionality of the problem (number
of parameters to sample) or the evaluation of the complex physical
forward model (nonlinear problems) in each MCMC iteration
[\citet{tarantola2004,buithanh2011,martinetal2012}].

Consider the model $\mathbf{Y} = f(\bolds{\beta}) + \bolds{\varepsilon}$ with
the physical forward model $f(\cdot)$, high-dimensional parameter
$\bolds{\beta}$ and error $\bolds{\varepsilon}$. In general, if the
physical problem is linear $(f(\bolds{\beta}) = X\bolds{\beta})$ and the
full conditional of $\bolds{\beta}$ is Gaussian, efficient sampling
from the high-dimensional Gaussian conditional distribution is
essential for the exploration of model space. In this case the error
$\bolds{\varepsilon}$ need not necessarily be Gaussian, but may be $t$ or
skewed-$t$ distributed [\citet{sahu2003,fruehwirth-schnatter2010}],
or a Gaussian error with a spatial correlation such as considered in
\citet{banerjee2003}. Given a sparse posterior precision matrix
[e.g., (\ref{betafullcond})], efficient sampling from a multivariate
normal can be carried out by Cholesky decomposition of a permuted
precision matrix as discussed in \citet{wilkinson2002} or
\citet{rue2005}, by using an approximate minimum-degree ordering
algorithm. A further improvement to the current sampling approach
might be to apply the Krylov subspace method from
\citet{simpsonetal2008}. This would require substantial
implementation efforts and is the subject of further research. If the
forward matrix or the prior precision matrix is not sparse, a dense
posterior precision matrix for $\bolds{\beta}$ will result. In this
case our sampling scheme is inefficient, but the model-space
reduction method developed by \citet{flathetal2011} might be used
instead. They exploit the low-rank structure of the preconditioned
Hessian matrix of the data misfit, involving eigenvalue
calculations. However, this approximation quantifies uncertainty of
large-scale linear inverse problems only for known hyperparameters,
thus ignoring uncertainty in those parameters. Eigenvalue
calculation in each MCMC step can be time consuming and prohibitive
for hierarchical models with unknown hyperparameters when the
posterior covariance matrix in every MCMC step changes. Here
additional research is needed.

% Extension of the current approach: 1) linear, non-Gaussian, fat tailed
If the full conditionals cannot be written as Gaussian [this case
includes the cases of a nonlinear $f(\cdot)$, a non-Gaussian prior
of $\bolds{\beta}$ or non-Gaussian, nonelliptical distributed
errors], using the standard MH algorithm to sample from the
high-dimensional posterior distribution is often computationally
infeasible. Constructing proposal density that provides a good
approximation of the stationary distribution while keeping the
high-dimensional forward model $f(\cdot)$ inexpensive to evaluate
has been the focus of the research over the past years:
\citet{liebermannetal2010} have drawn samples from an approximate
posterior density on a reduced parameter space using a
projection-based reduced-order model. In the adaptive rejection
sampling technique by \citet{cuietal2011}, the exact posterior
density is evaluated only if its approximation is accepted. The
stochastic Newton approach proposed by \citet{martinetal2012}
approximates the posterior density by local Hessian information,
thus resulting in an improvement of the Langevin MCMC by
\citet{stramertweedie1999}. Other random-walk-free,
optimization-based MCMC techniques for improving the proposal and
reducing correlation between parameters have been developed, such as
Hamiltonian Monte Carlo (HMC) [\citet{neal2010}], Adaptive Monte
Carlo (AM) [\citet{haarioetal2001,andrieuthoms2008}] and several
variations, for example, delay rejection AM (DRAM)
[\citet{haarioetal2006}], differential evolution MC (DEMC)
[\citet{terbraak2006}] and differential evolution adaptive Metropolis
(DREAM) [\citet{vrugtetal2009}], just to mention a few. However, MCMC
sampling of high-dimensional problems still requires a massive amount
of computing time and resources. For example, the quasi
three-dimensional nonlinear model of \citet{herbeietal2008}
contains about 9000 parameters on a $37\times19$ grid. We expect a long
computing time since they use standard MCMC sampling methods. The
example by \citet{cuietal2011} shows that their algorithm achieves
a significant improvement in both computing time and efficiency of
parameter space sampling for a large nonlinear system of PDEs that
includes about 10,000 parameters. However, their algorithm gives
11,200 iterations in about 40 days, while our problem requires only
38 hours (on a 32-core cluster) for the same number of iterations
for about 11,000 parameters.

While the future may be in effective uncertainty quantification of
nonlinear physical problems using model reduction and optimization
techniques, the computing time and resources at the moment are too
demanding to explore the large model space. This paper demonstrates
effective Bayesian analysis tailored to a realistically large
seismic tomographic problem, featuring over 11,000 structural and
source parameters. We deliver a precise uncertainty quantification
of tomographic models in terms of posterior distribution and
credible intervals using the MCMC samples, which allows us to detect
regions that differ from the reference earth model with high
posterior probability. Our approach is the first to solve seismic
tomographic problems in such high dimensions on a fine grid, and
thus provides ground work in this important research area.

\section*{Acknowledgments}

The authors acknowledge two referees, the Associate Editor and the
Editor for helpful remarks and suggestions which led to a significant
improved manuscript. The authors would like to thank the support of the
Leibniz-Rechenzentrum in Garching, Germany.

%suskaldyti doi

% imsref loaded by lrinkeviciute, 2013-04-17 08:52:10
% imsref loaded by lrinkeviciute, 2013-04-17 08:54:49

\printaddresses

\end{document}